\documentclass[sigconf,table]{acmart}
\usepackage[shortcuts]{extdash}
\usepackage{xcolor}
\usepackage{arydshln}
\usepackage{subfig}
\usepackage{graphicx}
\usepackage{enumitem}
\usepackage{xspace}
\usepackage{todonotes}
\usepackage{indentfirst}
\usepackage{ifthen}
\usepackage{hyperref}
\hypersetup{
	colorlinks   = true,
	citecolor    = blue,
	linkcolor    = blue,
	urlcolor=blue
}
\usepackage{cleveref}
\usepackage{graphics}
\usepackage{multirow}
\usepackage{makecell}
\usepackage{colortbl}
\usepackage{tcolorbox}
\usepackage[shortcuts]{extdash}

\AtBeginDocument{%
  \providecommand\BibTeX{{%
    \normalfont B\kern-0.5em{\scshape i\kern-0.25em b}\kern-0.8em\TeX}}}

\setcopyright{none}
\renewcommand\footnotetextcopyrightpermission[1]{}
\begin{document}

\title[Supporting Developers in Vulnerability Detection during Code
Review]{Less is More: Supporting Developers in\\ Vulnerability Detection during Code Review}

\author{Larissa Braz}
\email{larissa@ifi.uzh.ch}
\affiliation{%
 \institution{University of Zurich}
 \country{}
}

\author{Christian Aeberhard}
\email{christian.aeberhard2@uzh.ch}
\affiliation{%
	\institution{University of Zurich}
	\country{}
}

\author{Gül Çalikli}
\email{handangul.calikli@glasgow.ac.uk}
\affiliation{%
	\institution{University of Glasgow}
	\country{}
}

\author{Alberto Bacchelli}
\email{bacchelli@ifi.uzh.ch}
\affiliation{%
	\institution{University of Zurich}
	\country{}
}


\newcommand{\mindset}{mental attitude\xspace}
\newcommand{\mindsetB}{Mental attitude\xspace}

\newcommand{\dcs}{\textbf{DCS}\xspace}

\newcommand{\etal}{\textit{et al.}\xspace}
\newcommand{\eg}{\textit{e.g.,}\xspace}
\newcommand{\ie}{\textit{i.e.,}\xspace}

\newcommand{\rqOne}{Does explicitly asking developers to focus on security issues facilitate vulnerability detection during code review?\xspace}
\newcommand{\rqTwo}{How does the presence of a security checklist affect vulnerability detection during a security-focused code review?\xspace}
\newcommand{\rqTwoOne}{To what extent does the presence of a security checklist
affect vulnerability detection during a security-focused code review?\xspace}
\newcommand{\rqTwoTwo}{To what extent does the presence of a \emph{tailored}
security checklist affect vulnerability detection during a security-focused code review?\xspace}

\newcommand{\HNullOne}{The presence of security instructions does not facilitate vulnerability detection during code review.\xspace}
\newcommand{\HNullTwoOne}{The presence of a security checklist does not affect
vulnerability detection during a security-focused code review compared to providing security instructions.\xspace}
\newcommand{\HNullTwoTwo}{The presence of a \emph{tailored} security checklist
does not affect vulnerability detection during a security-focused code review compared to providing security instructions or a more generic checklist.\xspace}

\newcommand{\cbr}{{CBR}\xspace}
\newcommand{\ahr}{{AHR}\xspace}

\newcommand{\mdFive}{$MD5$\xspace}

\newcommand{\ncShort}{\textbf{SI}\xspace}  
\newcommand{\lcShort}{\textbf{SC}\xspace}  
\newcommand{\scShort}{\textbf{TC}\xspace} 
\newcommand{\nfShort}{\textbf{NI}\xspace} 

\newcommand{\ncFull}{Security Instructions\xspace}
\newcommand{\lcFull}{Security Checklist\xspace}
\newcommand{\scFull}{Tailored security Checklist\xspace}
\newcommand{\nfFull}{No security Instructions\xspace}

\newcommand{\ncShortNotBold}{SI\xspace}
\newcommand{\lcShortNotBold}{SC\xspace}
\newcommand{\scShortNotBold}{TC\xspace}
\newcommand{\nfShortNotBold}{NI\xspace}

\newcommand{\toolSteps}{five\xspace}
\newcommand{\totalPilot}{13\xspace}

\newcommand{\totalProfExperience}{\textcolor{red}{XXX}\xspace}

\newcommand{\itemsShortChecklist}{seven\xspace}
\newcommand{\itemsShortChecklistNumber}{7\xspace}
\newcommand{\cateShortChecklist}{three\xspace}

\newcommand{\itemsLongChecklist}{22\xspace}
\newcommand{\cateLongChecklist}{three\xspace}
\newcommand{\cateLongChecklistNumber}{3\xspace}
\newcommand{\subCateLongChecklistNumber}{7\xspace}
\newcommand{\itemsOnlyLongChecklist}{15\xspace}

\newcommand{\numberOfParticipantsExcludedStepOne}{357\xspace}
\newcommand{\numberOfParticipantsExcludedStepTwo}{15\xspace}
\newcommand{\numberAccesses}{522\xspace}
\newcommand{\numberFinished}{165\xspace}

\newcommand{\SDEshort}{\textbf{MSI}\xspace}
\newcommand{\CKVshort}{\textbf{BRA}\xspace}

\newcommand{\sde}{\emph{Sensitive Data Exposure}\xspace}
\newcommand{\sdeS}{\emph{SDE}\xspace}

\newcommand{\SDEfull}{\emph{Generation of Error Message Containing Sensitive Information}\xspace}
\newcommand{\CKVfull}{\emph{Use of a Broken or Risky Cryptographic Algorithm}\xspace}

\newcommand{\SDEshortNotBold}{MSI\xspace}
\newcommand{\CKVshortNotBold}{BRA\xspace}

\definecolor{gray50}{gray}{.5}
\definecolor{gray40}{gray}{.6}
\definecolor{gray30}{gray}{.7}
\definecolor{gray20}{gray}{.8}
\definecolor{gray10}{gray}{.9}
\definecolor{gray05}{gray}{.95}

\definecolor{purple}{HTML}{DADAEB}

\newlength\Linewidth
\def\findlength{\setlength\Linewidth\linewidth
	\addtolength\Linewidth{-4\fboxrule}
	\addtolength\Linewidth{-3\fboxsep}
}

\newenvironment{rqbox}{\par\begingroup
	\setlength{\fboxsep}{5pt}\findlength
	\setbox0=\vbox\bgroup\noindent
	\hsize=0.95\linewidth
	\begin{minipage}{0.95\linewidth}\normalsize}
	{\end{minipage}\egroup
	\textcolor{gray20}{\fboxsep1.5pt\fbox
		{\fboxsep5pt\colorbox{purple}{\normalcolor\box0}}}
	\endgroup\par\noindent
	\normalcolor\ignorespacesafterend}
\let\Examplebox\examplebox
\let\endExamplebox\endexamplebox

\definecolor{light-purple}{HTML}{dadaeb}
\newcommand{\rb}[1]{
	\begin{tcolorbox}[colback=purple,
		colframe=black,
		width=\columnwidth,
		arc=3mm, auto outer arc,
		boxrule=0.5pt,
		]
		#1
	\end{tcolorbox}
	\vspace{0.1cm}
}

\newcounter{Finding}
\stepcounter{Finding}

\newcommand{\roundedbox}[1]{
	\rb{
		\noindent
		\textit{\textbf{Finding \theFinding}. #1}
	}
	\stepcounter{Finding}
}


\newboolean{showcomments}
\setboolean{showcomments}{true}
	
\ifthenelse{\boolean{showcomments}}
{
	\newcommand{\nb}[3]{
		{\colorbox{#2}{\bfseries\sffamily\scriptsize\textcolor{white}{#1}}}
		{\textcolor{#2}{\textsf\small$\blacktriangleright$\textit{#3}$\blacktriangleleft$}}}
	 \newcommand{\version}{\emph{\scriptsize$-$Id$-$}}
}{
	\newcommand{\nb}[3]{}
} 

\newcommand{\hide}[1]{}

\newcommand{\ab}[1]{\nb{Alberto}{pink}{#1}}
\newcommand{\lari}[1]{\textcolor{magenta}{#1}}
\newcommand{\gul}[1]{\textcolor{red}{#1}}
\newcommand{\ca}[1]{\nb{Christian}{cyan}{#1}}

\newcommand{\vf}{VulnFound\xspace}    
\newcommand{\vnf}{VulnNotFound\xspace} 
\newcommand{\vfl}{VulnFound2\xspace} 
\newcommand{\vnfl}{VulnNotFoundLater\xspace}  
\newcommand{\reviewGroup}{Treatment\xspace} 

\newcommand{\familiar}{Familiarity\xspace}  

\newcommand{\incidents}{Incidents\xspace}  
\newcommand{\vulexp}{VulnExp\xspace}  
\newcommand{\intvuln}{IntroducedVuln\xspace} 
\newcommand{\foundvuln}{FoundVuln\xspace} 
\newcommand{\fixvuln}{FixedVuln\xspace} 
\newcommand{\exploitvuln}{ExploitedVuln\xspace} 
\newcommand{\practice}{UseInPractice\xspace} 

\newcommand{\courses}{Courses\xspace} 
\newcommand{\training}{TrainingType\xspace} 
\newcommand{\lec}{Lectures\xspace}  
\newcommand{\conf}{Conferences\xspace}  
\newcommand{\sem}{Seminars\xspace}  
\newcommand{\handson}{HandsOn\xspace}  
\newcommand{\proftrain}{ProfTraining\xspace}  
\newcommand{\update}{Update\xspace}  

\newcommand{\static}{StaticAnalysis\xspace} 
\newcommand{\dynamic}{DynamicAnalysis\xspace} 
\newcommand{\manual}{ManualAnalysis\xspace} 

\newcommand{\responsible}{Responsibility\xspace} 
\newcommand{\bug}{BugFound\xspace}  

\newcommand{\design}{Designing\xspace} 
\newcommand{\code}{Coding\xspace} 
\newcommand{\review}{Reviewing\xspace} 

\newcommand{\aware}{Awareness\xspace} 
\newcommand{\expert}{Expertise\xspace} 
\newcommand{\tools}{ToolUsage\xspace} 
\newcommand{\library}{ThirdPartyLib\xspace} 
\newcommand{\CRuse}{CRusage\xspace} 
\newcommand{\timepressure}{EnoughTime\xspace} 

\newcommand{\docTr}{DocAndTraining\xspace}  

\newcommand{\gender}{Gender\xspace}
\newcommand{\loe}{LevelOfEducation\xspace}
\newcommand{\emp}{EmploymentStatus\xspace} 
\newcommand{\role}{Role\xspace} 
\newcommand{\companysize}{CompanySize\xspace}
\newcommand{\teamsize}{TeamSize\xspace}
\newcommand{\oss}{OSSDev\xspace} 
\newcommand{\pde}{ProfDevExp\xspace} 
\newcommand{\je}{JavaExp\xspace}
\newcommand{\rp}{ReviewPractice\xspace}
\newcommand{\re}{ReviewExp\xspace}
\newcommand{\web}{WebDevExp\xspace}
\newcommand{\db}{DBDevExp\xspace}
\newcommand{\ce}{ChecklistExp\xspace}
\newcommand{\designFreq}{DesignFreq\xspace}
\newcommand{\codingFreq}{DevFreq\xspace}
\newcommand{\crFreq}{CRFreq\xspace}
\newcommand{\inter}{Interruptions\xspace}
\newcommand{\interfirst}{InterruptionsFirst\xspace}
\newcommand{\intersecond}{InterruptionsNext\xspace}
\newcommand{\tdexp}{DurationExp\xspace}
\newcommand{\tdrevisit}{DurationRevisit\xspace}
\newcommand{\reviewDuration}{DurationReview\xspace}

\newcommand{\pvalue}{\emph{p}\xspace}

\newcommand{\numParticipants}{150\xspace}
\newcommand{\numNC}{41\xspace}
\newcommand{\numNF}{33\xspace}
\newcommand{\numSC}{41\xspace}
\newcommand{\numLC}{35\xspace}
\newcommand{\numSecondModel}{117\xspace}
\newcommand{\totalStudents}{29\xspace}
\newcommand{\perStudents}{19\%\xspace}
\newcommand{\totalIToperators}{3\xspace}
\newcommand{\totalProf}{8\xspace}
\newcommand{\totalOtherJob}{9\xspace}
\newcommand{\totalManager}{6\xspace}
\newcommand{\totalEngineers}{93\xspace}
\newcommand{\perEngineers}{62\%\xspace}
\newcommand{\totalExperienceOverThree}{106\xspace}
\newcommand{\perExperienceOverThree}{71\%\xspace}
\newcommand{\totalExperienceOverTwo}{123\xspace}
\newcommand{\perExperienceOverTwo}{82\%\xspace}
\newcommand{\totalExperienceThreetoFive}{30\xspace}
\newcommand{\perExperienceThreetoFive}{20\%\xspace}
\newcommand{\totalExperienceSixtoTen}{31\xspace}
\newcommand{\perExperienceSixtoTen}{21\%\xspace}
\newcommand{\totalExperienceOverEleven}{45\xspace}
\newcommand{\perExperienceOverEleven}{30\%\xspace}
\newcommand{\totalOftenCR}{97\xspace}
\newcommand{\perOftenCR}{65\%\xspace}
\newcommand{\totalOftenDesign}{56\xspace}
\newcommand{\perOftenDesign}{37\%\xspace}
\newcommand{\totalOftenProgramming}{97\xspace}
\newcommand{\perOftenProgramming}{65\%\xspace}
\newcommand{\totalMale}{121\xspace}
\newcommand{\totalFemale}{7\xspace}
\newcommand{\totalOtherGender}{22\xspace}
\newcommand{\totalSDEfound}{55\xspace}
\newcommand{\totalCKVfound}{80\xspace}
\newcommand{\totalSDEmissed}{95\xspace}
\newcommand{\totalCKVmissed}{70\xspace}
\newcommand{\perSDEfound}{37\%\xspace}
\newcommand{\perCKVfound}{53\%\xspace}
\newcommand{\perSDEmissed}{63\%\xspace}
\newcommand{\perCKVmissed}{47\%\xspace}
\newcommand{\totalFoundAtLeastOne}{92\xspace}
\newcommand{\totalLCSCNone}{21\xspace}
\newcommand{\totalLCSCNotBoth}{48\xspace}
\newcommand{\totalLCSC}{76\xspace}
\newcommand{\totalNFsdeFound}{1\xspace}
\newcommand{\perNFsdeFound}{3\%\xspace}
\newcommand{\totalNCsdeFound}{19\xspace}
\newcommand{\perNCsdeFound}{46\%\xspace}
\newcommand{\totalLCsdeFound}{16\xspace}
\newcommand{\perLCsdeFound}{46\%\xspace}
\newcommand{\totalSCsdeFound}{19\xspace}
\newcommand{\perSCsdeFound}{46\%\xspace}
\newcommand{\totalNFckvFound}{7\xspace}
\newcommand{\perNFckvFound}{21\%\xspace}
\newcommand{\totalNFckvMissed}{26\xspace}
\newcommand{\perNFckvMissed}{79\%\xspace}
\newcommand{\totalNCckvFound}{25\xspace}
\newcommand{\perNCckvFound}{61\%\xspace}
\newcommand{\totalLCckvFound}{20\xspace}
\newcommand{\perLCckvFound}{57\%\xspace}
\newcommand{\totalSCckvFound}{28\xspace}
\newcommand{\perSCckvFound}{68\%\xspace}
\newcommand{\oddsRatioNFvsNC}{8\xspace}
\newcommand{\oddsRatioNFvsNCTXT}{eight\xspace}
\newcommand{\oddsRatioNFvsLC}{8\xspace}
\newcommand{\oddsRatioNFvsLCTXT}{eight\xspace}
\newcommand{\oddsRatioNFvsSC}{9\xspace}
\newcommand{\oddsRatioNFvsSCTXT}{nine\xspace}
\newcommand{\totalCKVnotNF}{73\xspace}
\newcommand{\totalSDEnotNF}{54\xspace}
\newcommand{\perCKVnotNF}{62\%\xspace}
\newcommand{\perSDEnotNF}{46\%\xspace}

\begin{abstract}
Reviewing source code from a security perspective has proven to be a difficult task. Indeed, previous research has shown that developers often miss even popular and easy-to-detect vulnerabilities during code review. Initial evidence suggests that a significant cause may lie in the reviewers' \mindset and common practices. 

In this study, we investigate whether and how explicitly asking developers to focus on security during a code review affects the detection of vulnerabilities. Furthermore, we evaluate the effect of providing a security checklist to guide the security review. To this aim, we conduct an online experiment with \numParticipants participants, of which \perExperienceOverThree report to have three or more years of professional development experience. Our results show that simply asking reviewers to focus on security during the code review increases \oddsRatioNFvsNCTXT times the probability of vulnerability detection. The presence of a security checklist does not significantly improve the outcome further, even when the checklist is tailored to the change under review and the existing vulnerabilities in the change. These results provide evidence supporting the \mindset hypothesis and call for further work on security checklists' effectiveness and design.

Preprint: \url{https://arxiv.org/abs/2202.04586}

Data and materials: \url{https://doi.org/10.5281/zenodo.6026291}
\end{abstract}



\maketitle

\section{Introduction}
A \emph{vulnerability} is a ``flaw or weakness in a system's design, implementation, or operation and management that could be exploited to violate the system's security policy''~\cite{shirey2000internet}.
The later vulnerabilities are discovered in the software development cycle, the higher the associated fixing costs are~\cite{planning2002economic}. 
Therefore, to avoid vulnerabilities, organizations are shifting security to earlier stages of software development~\cite{gitlab-survey}. 
However, security experts have to motivate and convince developers of the importance of finding vulnerabilities~\cite{Thomas:2018}. Yet, where to locate security within an organization remains a challenge~\cite{Tahaei:2019}. For instance, a programmer working solo is likely to create avoidable security problems because they can naturally have only one point of view~\cite{Weir:2017}. A solution to avoid these issues can be adopting security practices during code review.

Code review is a widely agreed-on practice~\cite{Boehm:2001} recognized as a valuable
tool for reducing software defects and improving the quality of software projects~\cite{Ackerman:1984, Ackerman:1989,Bacchelli:2013}. 
Previous studies show that code review is also an important practice for detecting and fixing security bugs earlier~\cite{Thompson:2017,McGraw:2004} and has positive effects on secure software development~\cite{Meneely:2012,Shin:2011,Meneely:2010}. 
 However, adopting security practices requires a large amount of knowledge which takes time to learn, and it can be hard to motivate~\cite{Turpe:2016,Poller:2017}. %
 In fact, security issues (even popular ones, such as \emph{Sensitive Data Exposure}~\cite{sensitiveExposure}) still often reach production code, despite code review practices. So, \emph{how can we better support code reviewers in detecting vulnerabilities?}

In the study we present in this paper, we investigate three interventions that aim to tackle the problem by guiding the focus of the reviewer. One is based on the \emph{developer's \mindset hypothesis} and two are based on an additional \emph{security checklists hypothesis}.

Studies in the literature indicate developers' \mindset as a leading cause for the introduction of vulnerabilities in the code~\cite{woon:2007, Xie:2011, Naiakshina:2019}. Specifically, vulnerabilities may be introduced because developers do not consider security as their responsibility~\cite{Naiakshina:2019} or strongly rely on other project members, processes, and technologies~\cite{woon:2007}. 
A potential solution to resolve issues related to developers' \mindset is giving explicit instructions regarding security. Indeed, \citet{Naiakshina:2017, Naiakshina:2018} showed positive effects when explicitly instructing computer science students and freelance developers to implement \emph{secure} password storage during coding. Nevertheless, writing and reviewing code are different activities~\cite{hermans2021programmer}, with even cultural differences among teams~\cite{Bacchelli:2013}, therefore evidence about the former may not translate to the latter.
Some preliminary yet promising evidence exists that the developers' \mindset could play a role in the context of code review too~\cite{braz:2021}. Using a one-group pretest-posttest experimental design~\cite{cook2002experimental}, \citet{braz:2021} found that a significant number of developers who missed a popular vulnerability during a code review could find it when explicitly warned about its presence.

In the context of code inspections~\cite{fagan2002design}, the use of checklists to support developers has been extensively studied~\cite{Rong2012, oladele2014empirical, lanubile2000evaluating, porter1995comparing} with positive results~\cite{dunsmore2003development}.
The OWASP foundation~\cite{owasp} proposes a popular code review guide that contains a security checklist~\cite{owaspChecklist}. It comprises items that guide the developers in finding security issues during a review.
Despite the positive results of using checklists during code inspection and the efforts by the OWASP foundation, the effectiveness of checklists in contemporary code review practices and to support vulnerability detection has not been established yet.

In our study, we set up three interventions (treatments) that we compare among themselves and to a review baseline (control). The baseline (\nfFull--\nfShort) consists of asking developers to perform a code review without giving special instructions. The first treatment (\ncFull--\ncShort), based on the \mindset hypothesis, explicitly instructs developers to perform the review from a security perspective. The second (\lcFull--\lcShort) and third (\scFull--\scShort) treatments additionally ask developers to use a security checklist in their review. The \lcShort checklist is derived from OWASP's Code Review Checklist~\cite{owaspChecklist}, while the \scShort one is a shorter version tailored to the change and security issues at hand, which we created to remove confounding factors caused by the length of the normal checklist.

We implement our study as an online experiment. A total of \numParticipants valid participants completed it. Among our participants, \perEngineers (\totalEngineers) reported being software engineers,  \perExperienceOverThree (\totalExperienceOverThree) reported three years or more of professional development experience, and \perOftenCR (\totalOftenCR ) reported performing code reviews daily.

Our results support the \mindset hypothesis: Participants instructed to focus on security issues were \oddsRatioNFvsNCTXT times more likely to detect vulnerabilities. Our results also call for further work on security checklists' effectiveness and design as they do not increase the detection of vulnerabilities in a security-focused review. 

\vspace{-1em}
\section{Background and Related Work}
In this section, we briefly introduce the concepts of code inspections and modern code review (the context of our study). Then, we review the literature on explicitly asking developers to use secure coding practices as a way to overcome issues due to the \mindset and on checklists for secure software development and inspections/reviews. We also provide background on the two security vulnerabilities we focus on in our online experiment. 

\textbf{Software developers and security.} %
Security often fails because of the lack of usability: users either misunderstand the security implications of their actions or turn off security features to workaround usability problems~\cite{Balfanz:2004}.
Software developers need to design systems that are both usable and secure. Yet, they still need support to create secure applications before addressing usable security~\cite{Tahaei:2019}. 

Existing findings about developers' attitude towards security are not conclusive.
On the one hand, a number of studies~\cite{Xiao:2014,Poller:2017,Assal:2018} found that developers prioritize more-visible functional requirements or even easy\-/to\-/measure activities, such as closing bug tracking tickets, over security. On the other hand, \citet{Christakis:2016} reported that developers care more about security issues than other reliability issues. 
\citet{Smith:2018} advocate that static analysis tools detect vulnerabilities and help developers resolve those vulnerabilities. Previous studies have proposed and improved tooling support according to developers’ needs~\cite{Ayewah:2008,Ayewah:2008-2,Smith:2015}, but tools are still generally poorly adopted by developers~\cite{Tahaei:2019} as they are confusing for developers to use~\cite{Smith:2018}.

\textbf{Developer-Centred Security (DCS).} %
\dcs studies have addressed some of developers' needs and attempted to apply existing methodologies from Human Computer Interaction and to adopt well\-/established usable security measures to software development~\cite{Wurster:2008,Green:2016,Pieczul:2017}.
The systematic literature review by~\citet{Tahaei:2019} points out the lack of research on several aspects of \dcs. For instance, they reported only one study in the context of code review~\cite{edmundson:2013}.
%

\noindent\textbf{Code inspections and modern code review.} 
Peer code review is a manual inspection of source code by developers other than the author. In 1976, \citet{fagan1976design} formalized a highly structured process for code reviewing, which includes a synchronous inspection meeting--code inspections. Over the years, researchers conducted several empirical studies on code inspections~\cite{kollanus2009survey}.

Nowadays, most organizations adopt a more lightweight code review process to limit the inefficiencies of inspections: modern code review~\cite{Cohe2010a}. This form of review is the focus of our study. Modern code review is asynchronous, tool- and change-based~\cite{baum2016factors}, and widely used in practice nowadays across companies~\cite{Bacchelli:2013,sadowski2018modern} and community-driven projects~\cite{rigby2013convergent,rigby2014peer}.

Contemporary code review mitigates several issues of code inspections but also removes the structure and checks devised to keep the reliability of the process and its outcome. Researchers provided strong evidence that, as a result of this different approach to inspection, the outcome of contemporary code review is less predictable and past theories from code inspections do not transfer  seamlessly~\cite{mantyla2008types,Bacchelli:2013,baum2016factors}. Therefore, it is important to conduct studies specific to this different context.

\noindent\textbf{Explicitly asking for secure coding practices.}  
Naiakshina et al.~\cite{Naiakshina:2017, Naiakshina:2018} conducted two studies with 40 computer science students. Half the participants received a task description that did not mention security, while the other half were explicitly tasked with implementing a secure solution. None of the participants in the first group stored passwords securely; while twelve participants in the group asked to create a secure solution implemented some level of security. 
Moreover, \citet{Naiakshina:2019} performed a similar study with freelance developers and found significant positive effect of security prompting on the secure storage of passwords.

\citet{Weir:2016} interviewed twelve industry experts to investigate how to improve the security skills of mobile app developers. They found that many of the most effective techniques for finding
security issues are \textit{dialectic}, \ie the discovery of knowledge through one
person questioning another. Some of the techniques recommended were penetration testing, code review, pair programming, and a variety of code analysis tools.
On the contrary, \citet{edmundson:2013} stated that manual code review could be expensive and impractical due to the need for several reviewers to inspect at a piece of code to find a vulnerability.
Later, \citet{braz:2021} conducted a one-group pretest-posttest code review experiment with software developers. They found that several developers often miss a popular and easy-to-detect vulnerability when reviewing code (pretest); yet, when explicitly informed about the presence of a vulnerability in the change, a significant portion of the additional developers could identify it (posttest).
These studies provide initial evidence that a specific, focused instruction on security could steer the developers' attention, thus overcoming their common \mindset to not consider security aspects in review.

\noindent\textbf{Checklist-supported inspections and review.}
Checklists have been mostly investigated for code inspections.
Ad Hoc Reading (\ahr) and Checklist-Based Reading (\cbr) are the standard reading techniques adopted by during code inspections~\cite{laitenberger:2000}. 
In a study with undergraduate students, \citet{oladele2014empirical} found that \cbr is effective in finding more issues during inspections with a 50\% decrease in false positives compared to \ahr. On the other hand, \citet{akinola2009empirical} did not find any significant differences in the efficiency of \ahr and \cbr in their study conducted with students in a distributed environment. Similarly, the findings of \citet{lanubile2000evaluating} and \citet{porter1995comparing} showed no significant differences between \ahr and \cbr in terms of  the number of defects detected during software requirements inspections.

Studies on checklists for contemporary code review are fewer. \citet{Rong2012} conducted a semi-controlled experiment with students and found evidence that checklists can help in guiding them during code reviews.
In the context of education, \citet{chong2021assessing} found that students are able to anticipate potential defects and create a relatively good code review checklist, which can be used to find defects. Finally, \citet{gonccalves2020explicit} registered an experiment report with the goal of investigating whether review checklists and guidance improve code review performance.

\noindent\textbf{Checklists in software security.} Previous studies have addressed the design of checklists for security-related aspects, such as software security requirements and security life-cycle~\cite{gilliam2003software,alam2010software,garrison2006computer}. 
\citet{gilliam2003software} provided guidelines for creating a software security checklist, emphasizing the importance of verifying security requirements beforehand to create a viable checklist.  
\citet{garrison2006computer} suggested that security checklists are particularly useful in guiding non-security professionals through a security-oriented software development process.
OWASP~\cite{owasp} stated that organizations with a proper code review process integrated into the software development life-cycle produced remarkably better code from a security standpoint~\cite{owaspChecklist}. 
To help businesses strengthen their security, such organizations developed different secure coding practices, which are widely adopted globally~\cite{Smarjov:2020}. In fact, they proposed a code review guide containing a security checklist~\cite{owaspChecklist}.

\citet{Cruzes:2017} investigated secure coding checklists' usage. However, the authors disagree that developers should adopt this practice. \citet{Smarjov:2020} assessed the OWASP checklist's effectiveness on the number of vulnerabilities reported to the HackerOne platform~\cite{HackerOne}. They found a moderate connection between filling out the checklist during the development phase and the distribution of the vulnerabilities reported in HackerOne. The likelihood of HackerOne participants finding a new vulnerability is 2.92 times higher if they do not follow the checklist during the code development. To best of our knowledge, no study has been conducted to investigate the impact of security checklists for security-focused code review. We address this question by analyzing whether a security checklist supports developers in this task. 

\noindent\textbf{Software Vulnerabilities.} OWASP~\cite{owasp} and the Common Weakness Enumeration Project (CWE)~\cite{CWETop25} are well-established projects that provide guidance and best practices for organizations to avoid security issues. Our study investigates whether security checklists support developers in vulnerability detection during code reviews. Therefore, among the OWASP's Top 10 Web Application Security Risks list~\cite{owaspTop10}, we focus on the \sde (\sdeS -- CWE~1029)~\cite{CWESDE} group, which 
can be 
mapped to items in our checklist. \sdeS occurrences have increased significantly over the last few years~\cite{shu2015privacy}, leading to damaging leaks. 
For instance, the largest instance of \sdeS to date impacted 3 billion Yahoo!'s user accounts, leaking their credentials (\ie email addresses, passwords, and security questions and answers). Security experts noted that the majority of the passwords used a strong hashing algorithm, but many used the \mdFive algorithm, which can be broken rather quickly~\cite{YahooBreach}.

There are many reasons for \sdeS, such as the Missing Encryption of Sensitive Data (CWE-311)~\cite{missingEncrypt} and the ones covered in our experiments, namely \SDEfull (\SDEshort - CWE 209)~\cite{CWELogExample} and \CKVfull (\CKVshort - CWE 327)~\cite{CWEBrokenCrypto}. The former refers to software that generates an error message including sensitive information about its environment, users, or associated data; while the latter refers to the use of a not recommended algorithm that may allow attackers to compromise data that has been protected.

\begin{figure*} [t]
	\centering
	\includegraphics[width=0.9\textwidth, angle=0]{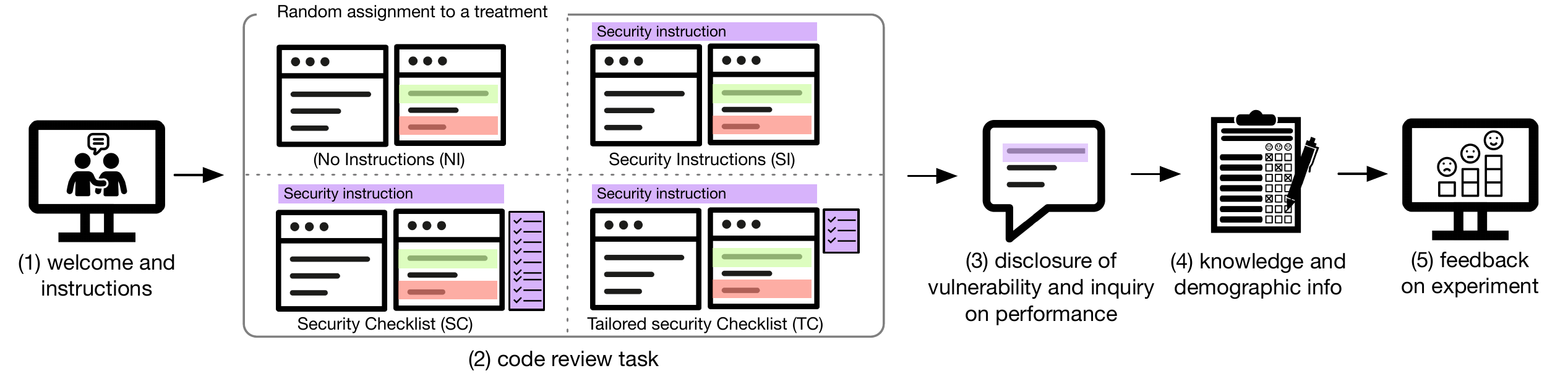}
	\caption{Steps of our online experiment.}
	\label{fig:experiment-flow}
\end{figure*}

\section{Research Methodology}

In this section, we introduce our research questions and detail our experimental design.

\subsection{Research Questions}

We structure our study around two research questions. We aim to understand the impact of (1) instructing developers to focus on security issues and (2) providing an additional checklists on vulnerability detection.

We formulate our first research question as follows:

\begin{center}
	\begin{rqbox}
		\begin{description}	    
			\item[]\textbf{RQ$_1$.} \emph{\rqOne}
		\end{description} 
	\end{rqbox}
\end{center} 

We hypothesize that explicitly asking developers to focus on security issues increases vulnerability detection during in code review, because it would steer developers' \mindset. Our formal hypothesis for \textbf{RQ$_1$} is: 

\smallskip
\begin{description}[leftmargin=0.3cm]
	\item[\textbf{H0$_{1}$:}]  \HNullOne	
\end{description}

Evidence shows that generic checklists aid developers during code inspections and can improve the inspections' outcome~\cite{oladele2014empirical, Rong2012}. 
We explore the hypothesis that providing developers with a security checklist (in addition to instructing them to do a security review) further increases the vulnerability detection during code review because it would guide the review tasks on relevant security aspects. We formulate our second research question as follows:

\smallskip
\begin{center}
	\begin{rqbox}
		\begin{description}	    
			\item[]\textbf{RQ$_2$.} \emph{\rqTwo}
		\end{description} 
	\end{rqbox}
\end{center} 

We organize \textbf{RQ$_2$.} in two sub-questions. First, we ask:

\begin{description}	    
	\item[]\textbf{RQ$_{2.1}$.} \rqTwoOne
\end{description}

The presence of items irrelevant to the change might deteriorate developers' code review performance. \citet{brykczynski1999survey} suggests that users are less likely to read through a multitude of checklist items. 
To mitigate this problem, one could imagine that future automation techniques will be able to generate a checklist \emph{tailored} to the code change to review. We manually create such an ideal checklist to measure its effect and ask:

\begin{description}	    
	\item[]\textbf{RQ$_{2.2}$.} \rqTwoTwo
\end{description} 

Our formal hypotheses for \textbf{RQ$_{2.1}$} and  \textbf{RQ$_{2.2}$} are as follows:

\begin{description}[leftmargin=0.3cm]
	\item[\textbf{H0$_{2.1}$:}]  \HNullTwoOne
	\item[\textbf{H0$_{2.2}$:}]  \HNullTwoTwo	
\end{description}

\subsection{Experimental Design}
\label{sec:design}

\begin{figure*} [t]
	\centering
	\includegraphics[width=0.95\textwidth]{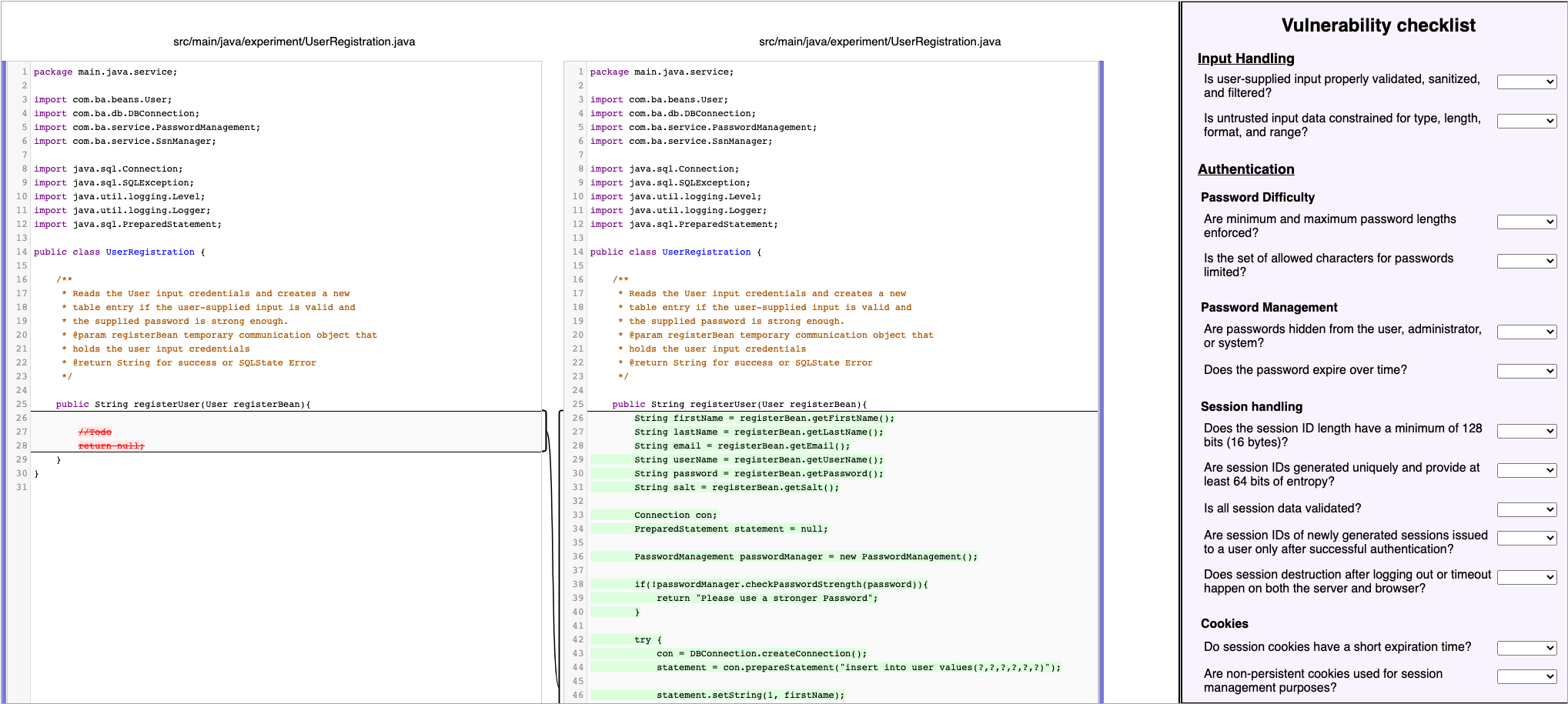}
	\caption{Example of the code review task with a security checklist using the tool.}
	\label{fig:tool-codereview}
\end{figure*}

\Cref{fig:experiment-flow} presents the flow of our online experiment, which consists of \toolSteps steps. 
Each step corresponds to one or two different web pages, and our experimental design does not allow the participants to return to previous pages or redo the experiment. 

\smallskip
\noindent\textbf{(1) Welcome page:} We provide the participants with general information about our study. We indicate that our goal, in general, is to improve code review practices and that they are going to do a code review. We also present our data handling policy to the participants and ask for their consent to use their data. 

\smallskip
\noindent\textbf{(2) Code review task:} We first ask the participants to take the review task as seriously as possible and to assume that the code they are going to see compiles and all tests pass. Depending on the treatment, participants may receive further instructions: 

\noindent\textbf{-- \nfFull} (\nfShort): Participants are provided no additional instruction.

\noindent\textbf{-- \ncFull} (\ncShort): Participants are informed that we are interested in security issues in th)is review. We also provide them with a definition of software vulnerability.

\noindent\textbf{-- \lcFull} (\lcShort): Participants receive the same instructions as in \ncShort. In addition, they are informed that they will have acces to a vulnerability checklist to assist them in the review and explain that each checklist item can be answered as yes/no/irrelevant with a specific checkbox. We ask them to work through every item in the checklist.

\noindent\textbf{-- \scFull} (\scShort): Participants receive the same instructions as in \lcShort but with a strictly tailored security checklist.

\noindent Finally, we provide all participants detailed instructions on how to add/edit/delete review comments.

When ready, the participants can press a button confirming that they have read all instructions and want to start the review. \Cref{fig:tool-codereview} shows a snapshot of the code review task that the participants assigned to \lcShort received after pressing the aforementioned button.

When the participants are done, they press a complete button. We warn participants assigned to \lcShort and \scShort about any unresolved checklist items, to ensure as high as possible checklist interaction. 
In addition, we ask participants of all four treatments whether they were interrupted during the code review and for how long. 

\smallskip
\noindent\textbf{(3) Disclosure of vulnerabilities and inquiry on performance:} After they submit their review, we inform participants that the source code contained two vulnerabilities (\SDEshort and \CKVshort) and show their location, types, definitions, and effects. Then, for each vulnerability we ask whether they found it. We ask participants to reason on why they could or could not detect it. 
In addition, we ask the participants assigned to the treatments \lcShort and \scShort whether the checklist helped them detect the vulnerabilities. We also ask their general feedback about the checklist, including whether and why they skipped any items. This step helps us better understand the impact of the security checklist on vulnerability detection and collect qualitative data for triangulating our quantitative findings.

\smallskip
\noindent\textbf{(4) Security knowledge and demographics:}
We ask a series of questions designed to gather information about factors that may affect the participants' detection of the vulnerabilities. Questions are mainly about the participants' security knowledge, checklist experience, and team culture. Most questions are in Likert scale format~\cite{Vagias:2006}. All questions are in the enclosed material~\cite{replication-package}. 
Finally, we ask a series of demographics questions about the participants' gender, highest education level, employment status, and years of experience in professional software development, Java programming, code reviewing, web application development, and databases. These questions are mandatory to fill in as collecting such data helps us identify which portion of the developer population our study participants represent~\cite{Falessi:2018}. We also ask about the frequency with which they designed, developed, and reviewed code in the last year; this helps us investigate possible confounding factors.

\subsection{Experimental Objects}
\label{sec:metho-objects}
The experimental objects consist of the code change to review, the injected vulnerabilities (\SDEshort and \CKVshort), and the digital checklists provided to participants assigned to the treatments \lcFull (\lcShort) and \scFull (\scShort). All the material is available in our replication package~\cite{replication-package}.

\smallskip
\noindent\textbf{Code Change.} The code change is implemented in Java: being Java one of the most popular programming languages~\cite{tiobe}, this allows us to reach a broader population of developers. The change comprises a feature implemented to manage users' online registration to a web service, consisting of two classes, five methods, and 160 lines. To avoid giving some participants advantages over others due to familiarity with the code, our code change does not belong to any existing codebase. Instead, we implemented a code change that is suitable for injecting both vulnerabilities. The code change is self-contained and suitable for being part of an existing software (\eg not a toy example to teach beginners Java programming). Finally, the code change is sufficient to consider possible attack scenarios, thus detecting the two vulnerabilities. 

We implemented the first version of the code change, after several brainstorming sessions. Later, we interviewed two security engineers with five and two years of professional security experience. They performed the experiment with treatment \lcShort. We asked them to read the code change before reading the checklist. Both security engineers pointed out the vulnerabilities. We also asked them to provide feedback on the checklist. Finally, we conducted a pilot study (\Cref{sec:pilots}). Based on the feedback we received from the security engineers and the participants of the pilot study, we iteratively modified the code change to ensure its realism and remove any confounding implementation or design-related issues other than the two vulnerabilities. 

\begin{figure}[t]%
	\centering
	\subfloat[\SDEfull (\SDEshort)]{%
		\includegraphics[clip,width=\columnwidth]{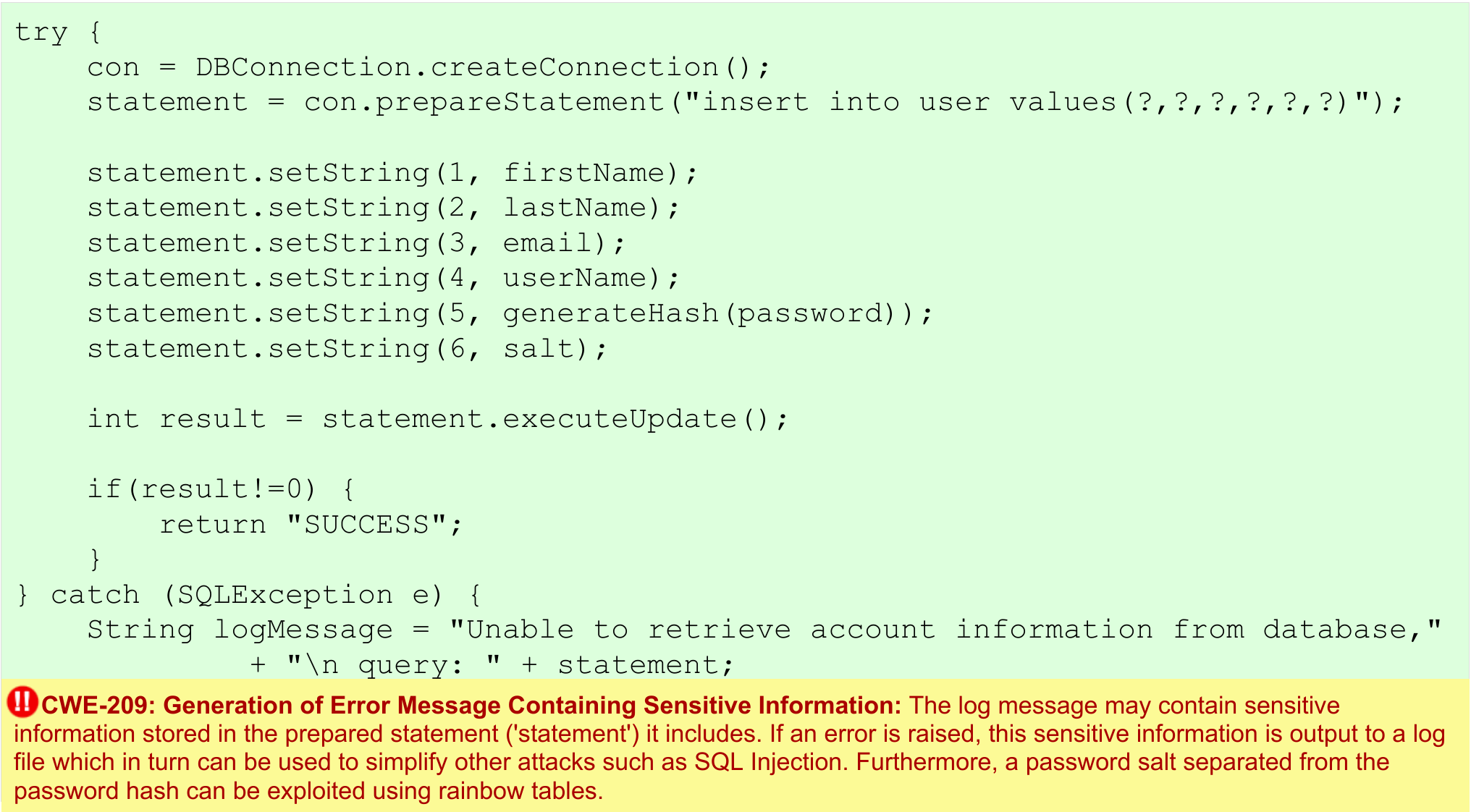}
	}
	
	\subfloat[\CKVfull (\CKVshort)]{%
		\includegraphics[clip,width=\columnwidth]{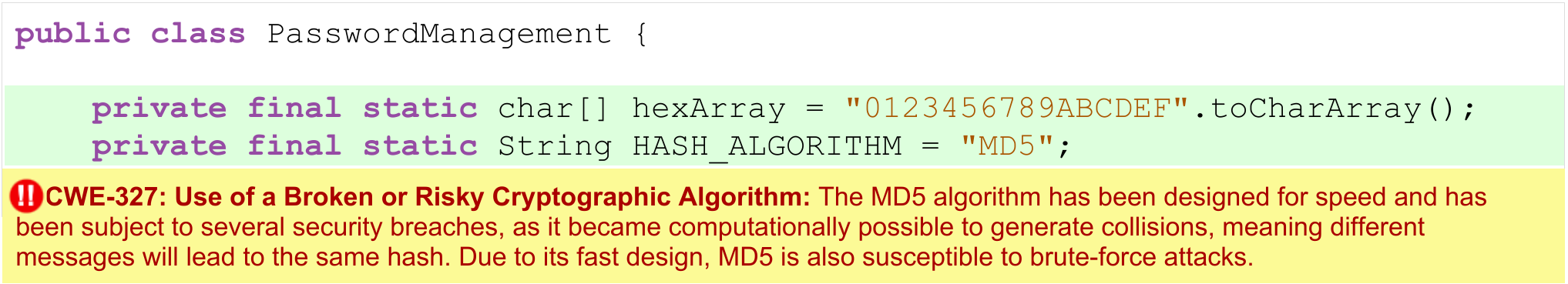}%
	}
	
	\caption{Object vulnerabilities used in our experiment.}
	\label{fig:vulnerabilities}
\end{figure}

\smallskip
\noindent\textbf{Security Vulnerabilities.} The code change contains two vulnerabilities. The first one is a \SDEfull (\SDEshort -- CWE~209)~\cite{CWELogExample}. When an \texttt{SQLException} is raised, the query containing an unencrypted plain text password and the user's email address is output to a log file. The error log will contain the user's sensitive information. The second vulnerability is the \CKVfull (\CKVshort --  CWE~327)~\cite{CWEBrokenCrypto}. The \emph{MD5} cryptographic hash function has been shown to be vulnerable to collision attacks. Different messages may have the same MD5 hash, making forgery attacks possible~\cite{sotirov2008md5}. \Cref{fig:vulnerabilities} shows the \SDEshort and the \CKVshort we used in our study.

We based our choice of vulnerabilities on the following criteria: prevalence, recognition, and discoverability using the selected checklist. Therefore, we selected vulnerabilities from the OWASP's ``Top 10 Web Application Security Risks'' list~\cite{owaspTop10} and we selected vulnerabilities that one could detect through the popular OWASP checklist~\cite{owaspChecklist} we employed (see more details below).
After the selection, we asked the two security experts to perform the review (we did not disclose the vulnerabilities). They were able to identify both vulnerabilities and pointed out which items in the checklist can help developers to detect these vulnerabilities. They further confirmed that these vulnerabilities are known and prevalent. 

\smallskip
\noindent\textbf{Security Checklists.}
The official checklist available in the OWASP Code Review Guideline has 78 items covering the most critical security controls and vulnerability areas (including \SDEshort and \CKVshort). We prepared the checklists for the \lcShort and \scShort treatments using the OWASP checklist and improved it by applying the guidance offered by the literature~\cite{brykczynski1999survey,zhu2016software, gilb1993software,chernak1996statistical}. The literature suggests that a checklist should be concise and no longer than one page for best practice and usability~\cite{gilb1993software,zhu2016software,brykczynski1999survey}; moreover, the categories, which represent particular features of the application, should guide the reviewers ``where to look" in the code~\cite{chernak1996statistical}. Therefore, we designed the checklist available to participants of \lcShort by reducing the OWASP checklist to \itemsLongChecklist security items structured in \cateLongChecklistNumber categories and \subCateLongChecklistNumber subcategories. 
The \lcShort checklist includes items relevant to the application context and the programming language, including two specific items related to the vulnerabilities in the code change. In addition, this checklist contains items irrelevant to the code change to represents a realistic situation in which not all the items are a perfect fit. Moreover, to facilitate participants' interaction with the checklist, we phrased each checklist item in the form of a question and provided a drop-down for the corresponding answer. Finally, the resulting checklist was used during our interview with two security engineers (\Cref{sec:metho-objects} -- \textbf{Code Change}). They provided feedback on each item of the checklist and pointed out the items that disclosed the vulnerabilities in the code change. To prepare the checklist for the treatment \scFull (\scShort), we reduced the checklist to a shorter version containing ideal-case scenario items, \ie only items relevant to the code change. This shorter checklist that we made available to \scShort participants contains \itemsShortChecklist security items structured in \cateShortChecklist categories.

\subsection{Variables, Measurements, and Analyses}
\label{sec:variables}


\begin{table}
    \caption{Variables used in the logistic regression models.}\label{tab:vars}
    \centering
    \begin{tabular}{l|l}
    \hline
    \cellcolor[HTML]{bcbddc}\textbf{Variable}             & \cellcolor[HTML]{bcbddc}\textbf{Description} \\ \hline\hline
    \multicolumn{2}{c}{\cellcolor[HTML]{dadaeb}\textit{Dependent Variables}}             \\ \hline
    \vf & \begin{tabular}[c]{@{}l@{}}The participant found the \\vulnerability in the code review\end{tabular}                \\\hline\hline  
     
    \multicolumn{2}{c}{\cellcolor[HTML]{dadaeb}\textit{Independent Variables}}             \\ \hline
    \reviewGroup                   & \begin{tabular}[c]{@{}l@{}} The treatment to which the\\ participant was assigned\end{tabular}       \\\hline\hline

    \multicolumn{2}{c}{\cellcolor[HTML]{dadaeb}\textit{Control Variables   (Review)}}             \\ \hline
    \inter               & \begin{tabular}[c]{@{}l@{}}For how long the participant was\\interrupted during the  review \end{tabular}      \\\hdashline[0.5pt/2pt]
    \reviewDuration                          & \begin{tabular}[c]{@{}l@{}}Duration of the code review\end{tabular}                                                           \\ \hline 
    \multicolumn{2}{c}{\cellcolor[HTML]{dadaeb}\textit{Control Variables   (Security Knowledge)}}             \\ \hline
    \familiar                      & \begin{tabular}[c]{@{}l@{}} Familiarity to vulnerabilities\end{tabular} \\ \hdashline[0.5pt/2pt]
    \courses                      & \begin{tabular}[c]{@{}l@{}} The participant has participated in \\security courses and/or training\end{tabular} \\ \hdashline[0.5pt/2pt]
    \update                      & \begin{tabular}[c]{@{}l@{}}  The participant keeps themselves up \\to date with security information \end{tabular}       \\\hline  
    \multicolumn{2}{c}{\cellcolor[HTML]{dadaeb}\textit{Control Variables   (Security Practice)}}             \\ \hline 
    \incidents                   & \begin{tabular}[c]{@{}l@{}} The participant has experience with \\ security incidents\end{tabular}       \\\hdashline[0.5pt/2pt]
    \responsible            & \begin{tabular}[c]{@{}l@{}}  The participant looks for vulnerabili- \\ties as a part of their job responsibility \end{tabular}       \\\hdashline[0.5pt/2pt]
     \multirow{3}{*}{\shortstack[l]{\{\design/\code/\\\review\}}}&The participant actively considers \\ & vulnerabilities 
     when \{designing \\ & software|coding|reviewing code\}    \\ \hline    
      \multicolumn{2}{c}{\cellcolor[HTML]{dadaeb}\textit{Control Variables (Demographics)}}             \\ \hline
    \loe                         & \begin{tabular}[c]{@{}l@{}}Highest achieved level of education\end{tabular}                                                          \\\hdashline[0.5pt/2pt]
    \emp                       & \begin{tabular}[c]{@{}l@{}}Employment status\end{tabular}                                                          \\\hdashline[0.5pt/2pt]
    \role                        & \begin{tabular}[c]{@{}l@{}}Role of the participant\end{tabular}                                                          \\\hdashline[0.5pt/2pt]
    \oss                         & \begin{tabular}[c]{@{}l@{}}The experience in OSS development \end{tabular}                                                          \\\hdashline[0.5pt/2pt]
    
    \multirow{5}{*}{\shortstack[l]{\{\pde |\\\je |\re |\\\web |\\\db | \\\ce\}}}&Years of experience \{as professional \\ 
    & developer | in java | in code review | \\
    & in web programming | in database \\
    & applications | using checklists \\ & during reviews\} \\\hdashline[0.5pt/2pt]
    
     \multirow{2}{*}{\shortstack[l]{\{\designFreq |\\\codingFreq | \crFreq\}}}&How often they \{design software | \\ & program | review code\} \\\hline 
    
    \end{tabular}
    \vspace{-1.5em}
    \end{table}

\Cref{tab:vars} presents all the variables we consider in our experiment. The independent variable ($\vf$) measures whether the participants found the vulnerability. The main independent variable is $\reviewGroup$ (\nfShort, \ncShort, \lcShort or \scShort). We consider the other variables as control variables, which also include the time spent on the review, the participant’s role, years of experience in java, code review, and using checklists during code reviews. Details about interruptions ($\inter$) are collected from the participants, and the duration ($\tdexp$) of each review is computed from the experiment’s log. To answer \textbf{RQ$_{1}$}, we build two multiple logistic regression models, one for each vulnerability (\SDEshort and \CKVshort) as dependent variable. The models are similar to the one used by \citet{Mcintosh:2016}, \citet{spadini2020primers}, and \citet{braz:2021}. 

To ensure that the selected logistic regression models are appropriate for the data we collect, we (i) reduced the number of variables by removing those with Spearman’s correlation higher than 0.5 using the VARCLUS procedure; (ii) ran a multilevel regression model to check whether there is a significant variance among reviewers, but we found little to none, thus indicating that a single level regression model is appropriate; and (iii) we added the independent variables into the model step-by-step and found that the coefficients remained stable.

\smallskip
\textbf{Analysis of code review outcome.} To give a value to our dependent variables (\ie $\vf$, whether the participants found the vulnerabilities), we do the following: (i) the first and last authors together inspect a subset of the remarks and checklist items interactions made by the participants during the review task and classify each vulnerability as detected or not; then, (ii) the first author classifies the remaining remarks and checklist items interactions; the first and last authors discuss the classification, especially unclear cases. The final decision is taken by cross-checking our classification with the answers participants gave when indicating whether they found the vulnerabilities (Step 3 in \Cref{fig:experiment-flow}).

\smallskip
\textbf{Analysis of open answers on performance.}
We use open card-sorting~\cite{spencer2009card} to analyze the the open answers participants' gave on their review performance (step 3, \Cref{fig:experiment-flow}). It allowed us to identify factors that might affect the detection of vulnerabilities during a review.
From the open-text answers, the first and second author separately created self-contained units, then sorted them into themes. To ensure the themes’ integrity, the authors iteratively sorted the units several times. Then, both authors compared and discussed their results to reach the final themes. 
The discussion helped to evaluate controversial answers, reduce bias caused by wrong interpretations of participants' comments, and strengthen the confidence in the outcome. 
We also use the card sorting output to triangulate our results and form new hypotheses, which we challenged with experimental data (\eg end of~\Cref{sec:results:rq1}).

\subsection{Pilot Runs}
\label{sec:pilots}

We conducted \totalPilot pilot runs to verify (1) the absence of technical errors in the experiment platform, (2) the ratio with which participants were able to find the injected vulnerabilities, (3) the understandability of the instructions and UI, (4) the absence of design/implementation issues beyond the injected vulnerabilities, and (5) the usability of the security checklist. We also gathered qualitative feedback from the participants. 

We conducted each run with a different participant. We recruited the participants through our professional network to ensure that they would take the task seriously and provide feedback on their experience. 
The participants’ data and qualitative feedback during the pilot runs were discussed iteratively among the authors every few pilot runs. We continued with our pilot iterations until the required changes were minimal. 
No data gathered from the pilot is considered in the final experiment.

\subsection{Recruiting Participants}

To recruit participants, the study authors spread the experiment through direct contacts from their professional network as well as their social media accounts, such as Twitter. In addition, the experiment was spread through practitioners blogs, web forums, and open source mailing lists. The actual aim of the experiment was not revealed. We introduced a donation-based incentive of 5 USD to a charity per complete and valid participant.


\section{Results}

\begin{figure}[t] 
	\centering
	\includegraphics[width=1\columnwidth,]{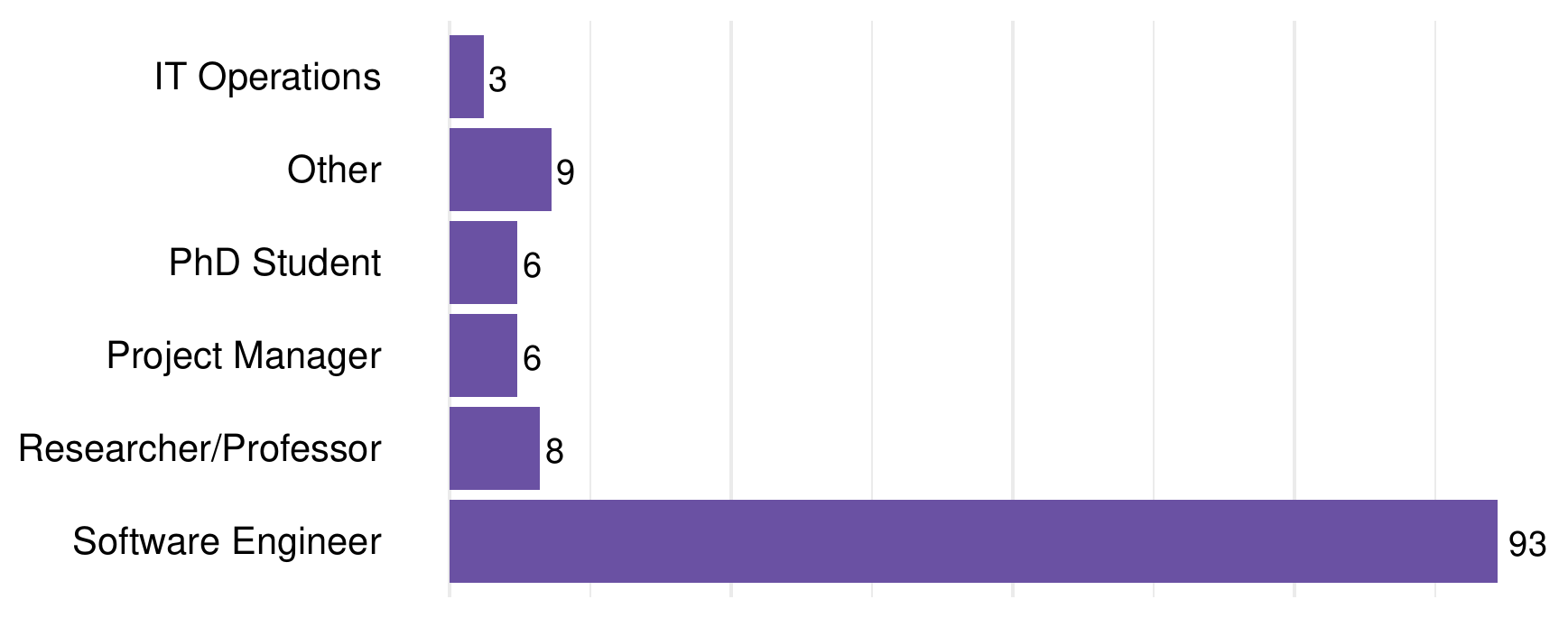}
	\caption{Job distribution among employed participants.}
	\label{fig:jobs}
\end{figure}

\begin{figure}[t]
	\includegraphics[width=1\columnwidth]{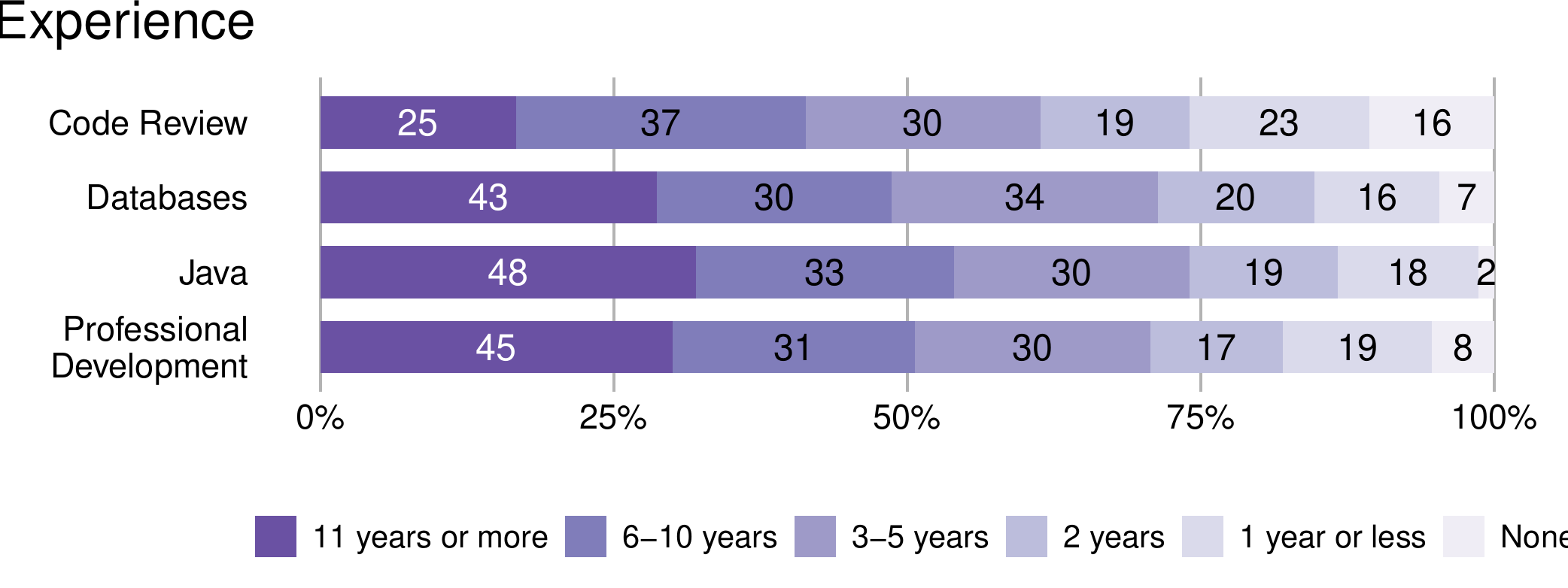}\\ 
	\vspace{0.25cm}
	\includegraphics[width=1\columnwidth]{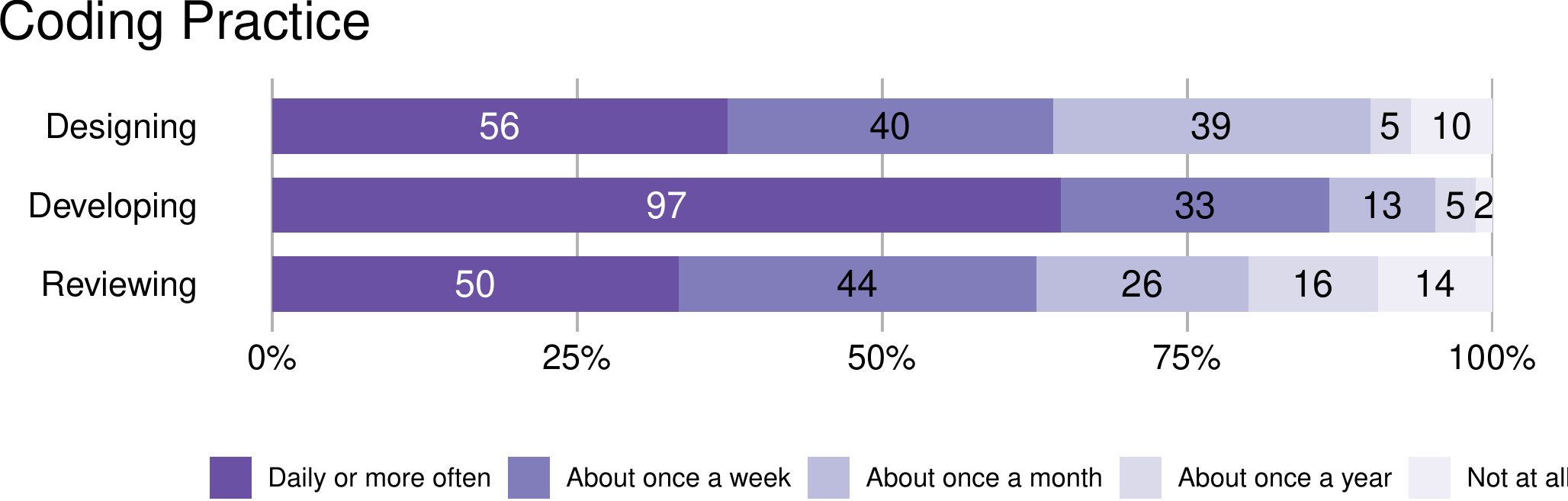}  
	\caption{Participants' demographics (absolute numbers).}
	\label{fig:demo}
\end{figure}
In this section, we report the results of our investigation.

\subsection{Validation of participants data}
A total of \numberAccesses people accessed our online experiment through the provided link. Of these, \numberOfParticipantsExcludedStepOne did not finish all experiment steps; thus, we removed their entries from the results dataset. We considered the \numberFinished people who completed all steps as potentially valid.
Then, we manually analyzed cases of participants whose code review duration was 1.5 times the interquartile range above the upper quartile or below the lower quartile. We removed participants who did not leave any remarks and did not interact with the qualitative questions of the performance inquiry. We also verified the checklist interactions (e.g., if they skipped all items of the checklist) of participants assigned to \scShort and \lcShort.
We removed \numberOfParticipantsExcludedStepTwo participants during this step. 
After applying these exclusion criteria, data from \numParticipants participants could be used for the analyses.

The valid participants were assigned to the treatments as follows: \numNF received \nfShort, \numNC received \ncShort, \numSC received \scShort, and \numLC received \lcShort. \Cref{fig:jobs} shows the current positions of participants with part-/full-time employment, and \Cref{fig:demo} presents the participants’ experience and practice.
Most participants currently have an engineering role (\perEngineers), have more than two years of professional development experience (\perExperienceOverTwo), and design, program, and review code daily (\perOftenDesign, \perOftenProgramming, and \perOftenCR, respectively).
\totalMale participants self-described as male, \totalFemale as female, and \totalOtherGender preferred not to disclose.

\subsection{RQ$_1$. Security instructions in review}
\label{sec:results:rq1}

\begin{table}[h]
	\centering
	\caption{Vulnerability detection by type and treatment.}
	\label{tab:rq1-results}
	\begin{tabular}{ccrr}
		\multicolumn{1}{l}{\textit{Treatment}} & \multicolumn{1}{l}{\cellcolor[HTML]{bcbddc}\textit{Vulnerability}} & \multicolumn{1}{c}{\cellcolor[HTML]{dadaeb}\textbf{Found}} & \multicolumn{1}{c}{\cellcolor[HTML]{dadaeb}\textbf{Not Found}}\\
		\toprule
		\multicolumn{1}{l}{\nfShort -- No security} & \SDEshortNotBold & 1 (3\%) & 32 (97\%) \\
		\multicolumn{1}{l}{Instructions} & \CKVshortNotBold & 7 (21\%) & 26 (79\%) \\
		\hline
		\multicolumn{1}{l}{\ncShort -- Security} & \SDEshortNotBold & 19 (46\%) & 22 (54\%) \\
		\multicolumn{1}{l}{Instructions} & \CKVshortNotBold & 25 (61\%) & 16 (39\%) \\
		\hline
		\multicolumn{1}{l}{\lcShort -- Security} & \SDEshortNotBold & 16 (46\%) & 19 (54\%) \\
		\multicolumn{1}{l}{Checklist} & \CKVshortNotBold & 20 (57\%) & 15 (43\%) \\
		\hline
		\multicolumn{1}{l}{\scShort -- Tailored} & \SDEshortNotBold & 19 (46\%) & 22 (54\%) \\
		\multicolumn{1}{l}{security Checklist} & \CKVshortNotBold & 28 (61\%) & 13 (39\%) \\
		\bottomrule
		\multicolumn{1}{l}{\multirow{2}{*}{\textbf{Total}}} & \SDEshortNotBold  & 55 (37\%) & 95 (63\%) \\
		\multicolumn{1}{l}{} & \CKVshortNotBold & 80 (53\%) & 70 (47\%) \\
	\end{tabular}
\end{table}

\Cref{tab:rq1-results} presents the results of the review in terms of vulnerability finding:
\totalFoundAtLeastOne participants found at least one of the vulnerabilities; \perSDEfound (\totalSDEfound) of the participants found the \SDEshort, while \perCKVfound (\totalCKVfound) found the \CKVshort. Among the control participants (\nfShort), \perNFsdeFound found the \SDEshort and \perNFckvFound found the \CKVshort; among the \numSecondModel participants in the treatments approximately \perSDEnotNF found the \SDEshort and \perCKVnotNF the \CKVshort.
When expressed in odds ratio, these results show that participants assigned to any of \ncShort, \lcShort, and \scShort are \oddsRatioNFvsNCTXT times more likely to find a vulnerability than those assigned to \nfShort ($p$ $<$ $0.001$).


\begin{table}[t]
\centering 
 \caption{Logistic regression models for \textbf{RQ$_1$}  (N=\numParticipants).} \label{tab:regression-rq1}
\begin{tabular}{lrrl|rrl}
& \multicolumn{3}{c|}{\cellcolor[HTML]{bcbddc}Dep. Var. = \SDEshort} & \multicolumn{3}{c}{\cellcolor[HTML]{bcbddc}Dep. Var. = \CKVshort} \\ 
 & \multicolumn{1}{c}{\cellcolor[HTML]{dadaeb}\textbf{Estim.}} & \multicolumn{1}{c}{\cellcolor[HTML]{dadaeb}\textbf{S.E.}} &  \multicolumn{1}{c|}{\cellcolor[HTML]{dadaeb}\textbf{Sig.}}  & \multicolumn{1}{c}{\cellcolor[HTML]{dadaeb}\textbf{Estim.}}  & \multicolumn{1}{c}{\cellcolor[HTML]{dadaeb}\textbf{S.E.}} &  \multicolumn{1}{c}{\cellcolor[HTML]{dadaeb}\textbf{Sig.}}  \\ 
  \hline
Intercept & -6.080 & 2.089 & ** & -5.893 & 1.741 & *** \\ 
  \lcShort & 3.809 & 1.165 & ** & 1.686 & 0.617 & ** \\ 
  \ncShort & 4.047 & 1.152 & *** & 1.627 & 0.617 & ** \\ 
  \scShort & 3.984 & 1.160 & *** & 1.933 & 0.631 & ** \\ 
  DurationCR & 0.040 & 0.013 & ** & 0.017 & 0.011 & \\ 
  Interruptions & -0.403 & 0.177 & * & -0.143 & 0.166 & \\ 
  Familiarity & 1.575 & 0.827 & . & 0.477 & 0.692 & \\ 
  Incidents & -0.703 & 0.602 &  & 0.332 & 0.504 & \\ 
  Practice & 0.566 & 0.498 &  & 0.672 & 0.462 & \\ 
  Update & -0.464 & 0.302 &  & 0.058 & 0.263 & \\ 
  Responsibility & 0.226 & 0.234 &  & 0.068 & 0.201 & \\ 
  Designing & -0.276 & 0.321 &  & - & - & -\\ 
  Coding & 0.815 & 0.326 & * & 0.192 & 0.223 & \\ 
  Reviewing & -0.159 & 0.240 &  & 0.176 & 0.207 & \\ 
  ProfDevExp & -0.017 & 0.186 &  & 0.176 & 0.207 & \\ 
  JavaExp & 0.110 & 0.194 &  & 0.173 & 0.169 & \\ 
  OftenDesign & -0.195 & 0.248 & & - & - & - \\ 
  DevFreq & -0.132 & 0.342 & & 0.180 & 0.254 &  \\ 
  CRFreq & 0.173 & 0.222 & & -0.106 & 0.198 &  \\ 
   \hline
\multicolumn{7}{r}{Sig. codes:  `***' $p <$ 0.001, `**' $p <$ 0.01, `*' $p <$ 0.05} \\ 
 \end{tabular}
\end{table}

In our logistic regression models, we set \nfShort as the reference level to better see the significance of the treatments (\ncShort, \lcShort and \scShort) in our \emph{Instructions} variable. We used the same starting variables in both models (see \Cref{sec:variables}), but the final ones differ due to removals during the multicollinearity analysis.
\Cref{tab:regression-rq1} shows the results of the logistic regression models considering as dependent variables whether the participants found \SDEshort and \CKVshort, respectively. The models confirm the result shown in \Cref{tab:rq1-results}: Instructing developers to focus on security and providing checklists is significant, thus, \emph{we can reject \textbf{H0$_1$}}.

\smallskip
\roundedbox{Developers who are instructed to focus on security issues during code review are \oddsRatioNFvsNCTXT times more likely to detect a vulnerability than participants who are not.}

\noindent\textbf{Qualitative Analysis -- \nfShort participants.} By analyzing the answers \nfShort participants gave on why they did (not) identify the vulnerability, we find recurring themes. Considering the case of \SDEshort, only one participant found it and explained: ``I work in a bank and in my company they send warning emails about this kind of sensitive information being logged (like card numbers being logged etc). So it caught my attention that what is being logged.''
The top-three reasons participants gave for not detecting \SDEshort are they (i) focused on aspects unrelated to security (nine mentions), overlooked the vulnerable code (six), or lacked the necessary knowledge (four). For instance, a participant reported: ``To be honest, I was optimizing for code structure/readability/architecture instead of security. Even so, I would probably have missed that one anyways.'' 

Concerning \CKVshort, \nfFull (\nfShort) participants reported finding it due to prior knowledge (four mentions) or the explicit use of the \mdFive algorithm (one). For instance, a participant wrote: ``It's very obvious that MD5 is being used to hash the password and MD5 being broken has been known for a long time now.'' Yet, \perNFckvMissed of the \nfShort participants did not find this vulnerability; the top-three reported reasons are: (i) lack of knowledge and experience (ten mentions); (ii) focus on aspects unrelated to security (five); and (iii) wrong assumptions about the code (four). Three participants reported that detecting the vulnerability was not part of their responsibility; as one put it: ``I am not very familiar with cryptography and I would assume there would be some separate security assessment.'' 

\noindent\textbf{Qualitative Analysis -- \ncShort participants.} The \ncShort participants received instructions to focus on security issues during the code review. Concerning \SDEshort they stated they found it because 
 (i) the code involves user or sensitive information that is valuable to hackers and raises security warning (eight mentions); (ii) of previous knowledge (one); and (iii) of first-hand experience (one). For example, a stated: ``It contains personal information which is to be protected according to the GDPR. The leak of the password hash could be a [vulnerability] depending on how [the access] to the logs differs from access to the database.'' 
The \ncShort who did not detect the \SDEshort explained that they missed it because they: (i) overlooked the vulnerability (six mentions), (ii) lacked knowledge or experience (four), and focused on factors unrelated to security (one). 

Regarding \CKVshort, the \ncShort participants mostly reported reasons related to previous knowledge as to why they found it: (i) prior knowledge or experience (seven mentions); the vulnerability is well-known (two); \mdFive can be hacked and quickly broken (two); and, \mdFive is an old vulnerability. A participant pointed out: ``It's rather well known that MD5 isn't secure anymore so I immediately noticed.' 
Among the \ncShort participants who did not find \CKVshort most (seven mentions) reported that they did not detect this vulnerability due to a lack of knowledge or experience. In addition, a participant stated to have made a wrong assumption: ``I was expecting that this could be configured in another part of the software.''

We challenged these qualitative reasons using data collected in step 5 (\Cref{fig:experiment-flow}). We used the variables described in \Cref{sec:variables} to map the reasons. We used Chi-Square test for the first two variables of $Knowledge$ ($Familiarity$ and $Courses$) and Mann-Whitney U test for $KnowledgeUpdate$ and all variables of $Practice$.  Regarding \SDEshort, only $Coding$ was significantly related to detecting this vulnerability ($p$ $<$ $0.01$). On the other hand, all variables of $Practice$ ($incidents$, $responsibility$, $design$, $coding$, $reviewing$) were significantly related to detecting \CKVshort (all with $p$ $<$ $0.01$). Additionally, $design$ from the $practice$ variables group was also significantly related ( $p$ $=$ $0.02$).

\roundedbox{Most developers perceive that security knowledge and practice influence their ability to detect vulnerabilities.}

\subsection{RQ$_2$. Security checklists in reviews}

\begin{table}[t]
\centering 
 \caption{Logistic regression models for \textbf{RQ$_2$} (N=\numSecondModel).} \label{tab:regression-rq2}
\begin{tabular}{lrrl|rrl}
	& \multicolumn{3}{c|}{\cellcolor[HTML]{bcbddc}Dep. Var. = \SDEshort} & \multicolumn{3}{c}{\cellcolor[HTML]{bcbddc}Dep. Var. = \CKVshort} \\ 
 & \multicolumn{1}{c}{\cellcolor[HTML]{dadaeb}\textbf{Estim.}} & \multicolumn{1}{c}{\cellcolor[HTML]{dadaeb}\textbf{S.E.}} &  \multicolumn{1}{c|}{\cellcolor[HTML]{dadaeb}\textbf{Sig.}}  & \multicolumn{1}{c}{\cellcolor[HTML]{dadaeb}\textbf{Estim.}}  & \multicolumn{1}{c}{\cellcolor[HTML]{dadaeb}\textbf{S.E.}} &  \multicolumn{1}{c}{\cellcolor[HTML]{dadaeb}\textbf{Sig.}}  \\ 
  \hline
  Intercept & -1.948 & 1.832 &  & -6.007 & 2.102 & ** \\ 
  \lcShort & 0.244 & 0.557 & & -0.016 & 0.571 & \\ 
  \scShort & 0.167 & 0.546 & & 0.288 & 0.569 & \\ 
  DurationCR & 0.0425 & 0.014 & ** & 0.023 & 0.013 & . \\ 
  Interruptions & -0.402 & 0.181 & * & -0.291 & 0.190 & \\ 
  Familiarity & 1.636 & 0.843 & . & 0.619 & 0.777 & \\ 
  Incidents & -0.535 & 0.629 & & 0.441 & 0.635 & \\ 
  Practice & 0.681 & 0.519 &  & 0.817 & 0.545 & \\ 
  Update & -0.505 & 0.309 & & 0.005 & 0.304 & \\ 
  Responsibility & 0.186 & 0.241 &  & 0.078 & 0.244 & \\ 
  Designing & -0.344 & 0.330 & & -0.033 & 0.324 &  \\ 
  Coding & 0.829 & 0.335 & * & 0.190 & 0.262 & \\ 
  Reviewing & -0.124 & 0.244 &  & 0.234 & 0.245 & \\ 
  ProfDevExp & -0.055 & 0.195 &  & 0.082 & 0.185 & \\ 
  JavaExp & 0.122 & 0.202 &  & 0.182 & 0.206 & \\ 
  OftenDesign & -0.130 & 0.251 &  & 0.022 & 0.254 & \\ 
  DevFreq & -0.189 & 0.355 &  & 0.630 & 0.365 & . \\ 
  CRFreq & 0.147 & 0.230 &  & -0.083 & 0.230 & \\ 
   \hline
\multicolumn{7}{r}{Sig. codes:  `***' $p <$ 0.001, `**' $p <$ 0.01, `*' $p <$ 0.05} \\ 
 \end{tabular}
\end{table}

In \textbf{RQ$_2$} we investigate the effect of providing a security checklist in addition to asking developers to focus on security issues. In this analysis, we thus remove the participants of \nfFull (\nfShort) and set treatment \ncFull (\ncShort) as the reference level for our logistic regression models. 

\Cref{tab:regression-rq2} shows the results. While $Duration$ and $Interruptions$ are still significant for the detection of \SDEshort, neither \scShort nor \lcShort treatments are significant. Therefore, \emph{we cannot reject \textbf{H0$_2$}}. The presence of a security checklist does not affect the detection of software vulnerabilities during code review compared to only asking developers to look for security issues. 

\roundedbox{Developers receiving a checklist (even if tailored to the code change) did not find more vulnerabilities than developers only instructed to focus on security issues.}

\noindent\textbf{Qualitative Analysis -- \scShort and \lcShort participants.} We asked \scShort and \lcShort participants whether the checklist helped them detecting the vulnerabilities. 

\textit{Checklist was helpful:} Concerning \SDEshort, the top-three reasons given by participants about why the checklist helped them finding it are: the checklist 
(i) marked the participants recheck the code (eight mentions),
(ii) pointed out specific areas (seven), and 
(iii) primed the participants to look for vulnerabilities (three). 
A participant explained: ``Did not see the data passed in the logger at first. But after seeing that item in the checklist I went back to check again.'' Two participants stated to not need the checklist to find the vulnerability but they still reported it as helpful; one stated: ``Even though I found this vulnerability before looking at the checklist, it made me specifically look for that again which is probably a good thing.''

Regarding \CKVshort, the top-three reported reasons for the checklist being helpful are that the checklist: (i) helped participants to look for vulnerabilities (five mentions), the checklist pointed out specific areas (five), and led them search for further information (external sources) regarding the vulnerability (three). 
A participant explained: ``Did not notice which algorithm was used at first. After seeing the checklist I remembered the issue about the md5 algorithm.' 'Another stated: ``While review, I search[ed on] Google how strong MD5 is.'' 

\textit{Checklist was not helpful:}
Some \scShort and \lcShort participants who identified the vulnerabilities stated that the checklist did not help. Regarding the \SDEshort, the top-four reported answers are: (i) the participant noticed the vulnerability before using the checklist (five mentions); (ii) the participant already looks for this vulnerability (one); and (iii) the vulnerability was easy to spot (one). For instance, a participant reported: ``I was already on the lookout for that kind of mistake, because in a past job part of my responsibility was making sure that didn't happen.''
Regarding the \CKVshort, the top-four reported answers are: (i) the participant already knew about this vulnerability (15 mentions), (ii) the checklist was not specific enough (four), (iii) the participant did not pay attention to the checklist (two), and this vulnerability requires explicit security knowledge (two).

Participants in \scShort and \lcShort who did not find the vulnerabilities also explained why the checklist did not help. Regarding \SDEshort, the top-four answers are: (i) lack of attention (three mentions); (ii) they focused on something unrelated to security (three); and (iii) they overlooked the vulnerability (two). For instance, a participant reported: ``I wasn't focused on the exception part, hence I missed the vulnerability.''
Regarding the \CKVshort, most participants reported lack of knowledge (six mentions) as the reason the checklist did not help them detect the vulnerability. A participant explained: ``I wasn't aware that MD5 was this vulnerable: I estimated during the review that it was a good choice.''

\roundedbox{Overall, 32 out of 76 of the developers who received a security checklist perceived it as helpful. Yet, ten reported to be able to detect the vulnerabilities without it.}

\subsection{Robustness Testing}
\label{sec:results:robustness}

We employ \textit{robustness testing}~\cite{neumayer:2017} to further challenge the validity of our findings. 

\smallskip
\noindent\textbf{Finding the first vulnerability distracted the participants.}
Participants who did not receive security instructions (treatment: \nfFull --\nfShort) might have stopped looking for the second vulnerability after they found the first one assuming they found the only problem. To challenge our results against this hypothesis, we simulated what the results would have been if all participants who detected one vulnerability instead detected both (\ie they did not stop after finding the first). In the simulation, eight \nfShort participants found both vulnerabilities. Using the simulated data, we re-run the analyses and checked the new results. Our logistic regression models achieved the same results as for the original data: the presence of security instructions (as well as the security checklists) is significant to the detection of both vulnerabilities. Additionally, $Coding$ also remains significant. Therefore, even if finding the first vulnerability distracted the participants, this did not impact the final results.

\smallskip
\noindent\textbf{Participants have different levels of experience/practices.} One factor that may impact the participants' performance in the review task is their experience and practice (\eg developers with fewer years of experience might find it more challenging to identify the vulnerability). It is important that the treatment groups are balanced in this aspect. In addition to adding experience and practice as control factors in our regression models, we also performed multiple pairwise-comparison between the means of the treatments using the Tukey HSD test and correcting for multiple tests with the Holm approach. We found no statistically significant difference.

\smallskip
\noindent\textbf{The number participants is too low.} 
To calculate the minimum size sample of our experiment (\ie number of participants with valid responses), we performed a preliminary power analysis using the G$^{\star}$Power~\cite{Faul:2007}, using \textit{Two-tail} test with \textit{odds ratio} $= 1.5$, $\alpha = 0.05$, \textit{Power} = $1 - \beta$ $= 0.95$, and $R^2$ $=0.3$. We used a manual distribution.
After running the analysis, we found that our experiment needed a minimum of 143 participants.
As our experiment reached \numParticipants valid participants (bigger than necessary), we believe that our sample is representative. 
However, this sample size is valid only for the first logistic regression model that we built to answer the research question \textbf{RQ$_{1}$}. 
To build the second logistic regression model for \textbf{RQ$_{2}$}, we exclude the participants assigned to \nfShort. Therefore, for this analysis, we reduced our participants' number to \numSecondModel, which is still a  quite large sample compared to many experiments in software engineering~\cite{Baum:2019}. However, we also conducted other statistical tests to verify the effect of single variables on the expected outcome and reported the results as the sample size could have affected the significativity of the multivariate statistics.

\smallskip
\noindent\textbf{The vulnerabilities are too easy/hard to detect.}
The vulnerabilities injected in the changes might have affected the validity of our results. Reviewers might not find a too hard to detect vulnerability and get discouraged to continue the experiment, while participants might detect a too easy vulnerability regardless of any other influencing factor, even without paying too much attention to the review. We measure that \perSDEfound and \perCKVfound of the participants found the \SDEshort and the \CKVshort, respectively, suggesting that these vulnerabilities were neither too trivial nor too difficult to find.

\section{Threats to validity}

\noindent\textbf{Internal Validity.} As our experiment was online, participants conducted it in different environments (\eg noise level and web searches), which could have influenced the results; yet it is expected that developers in real world settings also work with various tools and environments. Interruptions could have also influenced the results, so we asked participants about interruptions and their duration and included this information in our statistical analyses. To mitigate the possible threat from missing control over subjects, we included some questions to characterize our sample, such as experience and role (Step 4 in \Cref{fig:experiment-flow}). We conducted statistical tests to investigate how these factors affect the results.
Participants were allowed to take the experiment only once to prevent duplicate participation. We removed participants who did not complete the experiment and manually analyzed outliers in terms of duration.

We acknowledge that developers' background, knowledge, and practice may impact the experiment's results. However, we could only register a statistically significant (positive) effect for the variable `coding` (i.e., ``the participant considers vulnerabilities when coding'' -- see \cref{tab:vars}) when finding the \SDEshort vulnerability (Tables~\ref{tab:regression-rq1} and~\ref{tab:regression-rq2}). In their open-text responses, developers indicated that they perceive these aspects (especially knowledge) to play a substantial role. In addition, we performed multiple pairwise comparisons between the means of the treatments using the Tukey HSD test and correcting for multiple tests with the Holm approach (\cref{sec:results:robustness}). We believe the impact of background knowledge and practice should be investigated with further studies.

\smallskip
\noindent\textbf{Construct Validity.} The vulnerabilities were based on the examples of their CWE description~\cite{CWELogExample,CWEBrokenCrypto} and the security checklists were based on the OWASP code review checklist~\cite{owaspChecklist}. To mitigate the risk that the code changes and security checklists could have unanticipated characteristics that biased the results (\eg due to inappropriate quality), we validated them with two security experts (\Cref{sec:metho-objects}) and through \totalPilot pilot runs (\Cref{sec:pilots}).

The online platform showed all code on the same page, and participants had to scroll down to proceed to the next page of the online experiment. In this way, we aimed to ensure that subjects saw the complete code change.
To mitigate the fact that the online reviewing platform may differ from what participants are used to, the experiment tool's interface is designed to be identical to the popular review tool Gerrit~\cite{Gerrit}.
Documentation was also added to make the experiment closer to a real world scenario. 

We used different measurement techniques to mitigate \emph{mono-method bias}~\cite{Cook:1979}: we obtained qualitative results by employing card sorting on participants’ responses to their performance inquiry (Step 3 in \Cref{fig:experiment-flow}). We also used this technique on the feedback of \scShort and \lcShort participants on the helpfulness of the security checklist.
We ran statistical analyses on variables from participants’ answers (Step 4 in \Cref{fig:experiment-flow}) and triangulated the qualitative and statistical findings. 
To mitigate \emph{mono-operation bias}~\cite{Cook:1979}, we used more than one variable to measure each construct (\eg security knowledge, practice). Each of variable (see \Cref{tab:vars}) corresponds to a question in the survey on vulnerabilities (Step 4 in \Cref{fig:experiment-flow}). 
To mitigate \emph{interaction of different treatments}~\cite{Cook:1979}: participants were randomly assigned to one of the treatments (\nfShort, \ncShort, \scShort or \lcShort). To test \textbf{H0$_{1}$} (\ie the effect of a security instructions on vulnerabilities detection), we analyzed responses of all participants. To test \textbf{H0$_{2.1}$} (\ie the effect of a security checklist on vulnerability detection) and \textbf{H0$_{2.2}$} (i.e., the effect of a tailored security checklist on vulnerability detection), we considered responses of participants assigned to \ncShort, \scShort and \lcShort.

We used the qualitative data--collected from \lcShort and \scShort ($N=76$)--to understand developers’ perceptions of the checklist’s effect. We used the qualitative data from \nfShort and \ncShort participants ($N=74$) to get insights into their perceived reasons for (not) finding the vulnerabilities. Therefore, qualitative answers in these two groups (\lcShort-\scShort, \nfShort-\ncShort) refer to two different contexts. We used this design to limit the number of questions (especially those with open-text answers) to prevent overloading our participants and to ensure broader participation in the experiment.

\smallskip
\noindent\textbf{External Validity.} To have a diverse sample of participants, we invited software developers from several countries, organizations, education levels, and background; yet, our sample is not representative of all developers. Moreover, further studies should be designed and run to establish the generalizability of our findings with different changes and vulnerabilities. 

\section{Discussion}

We investigated the effects of instructing developers to focus on security issues and of two security checklists on vulnerability detection in code review. 
We now discuss the main implications and results of our study.

\smallskip
\noindent\textbf{\mindsetB  matters.} 
Vulnerabilities can lead to strong negative consequences when they go undetected. Almost every day the media publishes news about successful cyberattacks, showing more and more the urgency and need for advances in the secure software engineering field. 
Previous studies reported \mindset as one leading cause of why vulnerabilities are introduced in the code~\cite{woon:2007, Xie:2011, Naiakshina:2019} and initial evidence suggests that it may be one of the reasons  for not being detected during code review~\cite{braz:2021}.

In line with previous findings~\cite{braz:2021,woon:2007,Naiakshina:2019}, our results suggest that developers' \mindset plays an important role in the detection of software vulnerabilities during code reviews. 
These results strengthen the questions on the effectiveness of the current development process, including coding and reviewing activities. 
Organizations may consider incorporating Secure Code Review (SCR) into their development process to create a different approach. SCR is an enhancement to the standard code review practice where the structure of the review process places security considerations, such as company security standards, at the forefront of the decision-making~\cite{owaspChecklist}. Further studies need to be designed and carried out to determine how to better design SCR and evaluate its effectiveness.

Interestingly, our results show that security instructions are helpful but the detection of vulnerabilities does not increase when a security checklist is added to the code review. Thus, a \emph{less is more} approach can be adopted to improve the code review process: keeping it as simple as possible with security instructions rather than incorporating advanced or complicated solutions, such as security checklists. Studies can be carried out to investigate how to effectively address security instructions during SCR.

\smallskip
\noindent \textbf{Security checklists are no silver bullets.}
Checklists are a well\-/established reading support mechanism often used by individual inspectors for the purpose of preparation~\cite{Dunsmore:2003}. Previous studies~\cite{oladele2014empirical,Rong2012} found that Checklist-Based Reading effectively finds more issues during code reviews than Ad Hoc Reading. 
\citet{braz:2021} suggested security checklists as a way to improve the code review process. 
Surprisingly, we found that they do not facilitate the detection of software vulnerabilities compared to just instructing the developers to focus on security issues during the code review, even when the checklist is strictly tailored to the code change. 

Our findings provide initial evidence that reviewers may make wrong assumptions, such as assuming that the developer already implemented the change considering the application's security. 
In fact, checklists may not always be helpful. For example, it can remind the developer to use hashing to protect passwords. However, in practice, hashed passwords cannot be used for challenge-response authentication, negotiating a shared cryptographic session key, or for other such things~\cite{Bellovin:2008}. Thus, the developer must decide between protecting against a stolen password file or being able to generate a shared session key. Although well-intentioned, a checklist cannot answer this question and, without proper knowledge, developers may take the wrong decision. 
In our study, developers reported lack of knowledge as one of the main reasons why the checklists did not help in detecting the vulnerabilities. 
A security checklist could be used as a reminder but not as a substitute for analysis~\cite{Bellovin:2008}. 

A possible problem of checklists is that they are often too general and do not sufficiently tailored to a particular development environment~\cite{laitenberger:2000}. However, our results indicate that a security checklist strictly tailored to the code change does not increase the vulnerability detection in code reviews. This finding raises questions on whether the problem is not the security checklists but how developers perceive them. In fact, a few developers reported to have read the checklist \emph{after} the review or to not consider it at all. Further studies could investigate how to effectively integrate security checklist in tools and processes to support review.

\section{Conclusion}

We investigated whether and to what extent instructing developers to focus on security issues and providing security checklists during code reviews can support the detection of software vulnerabilities. Specifically, we designed and conducted an experiment with \numParticipants participants, which included a code review task and a survey. The code change to review contained two vulnerabilities: \SDEfull (\SDEshort) and \CKVfull (\CKVshort).
The participants were randomly assigned to one of the following treatments: \nfFull (\nfShort), \ncFull (\ncShort), \lcFull (\lcShort), or  \scFull (\scShort). The latter two groups received a security checklist; \ncShort received instructions to focus on security issues during the review; and \nfShort received neither.

In total, \totalFoundAtLeastOne participants found at least one of the vulnerabilities. Those who received the instructions to focus on security issues were at least \oddsRatioNFvsNCTXT times more likely to detect vulnerabilities. However---surprisingly---adding security checklists did not increase further the vulnerabilities detection during code review compared to only receiving security instructions.

Our findings provide evidence that developers' \mindset plays a role in detecting software vulnerabilities during code reviews. The effect of the security instructions provides evidence that vulnerability detection could be triggered with proper security considerations, such as security standards for code reviews. Moreover, the role and design of security checklists should be investigated further to establish and improve their effectiveness.

\begin{acks}
The authors would like to thank the anonymous reviewers for their thoughtful and important comments, which helped improving our paper. The authors also express gratitude to the \numParticipants valid participants who took part in the study, and gratefully acknowledge the support of the Swiss National Science Foundation through the SNF Projects No. PP00P2\_170529 and PZ00P2\_186090.
\end{acks}

\newpage

\bibliographystyle{ACM-Reference-Format}
\bibliography{icse22.bib}


\begin{thebibliography}{92}


\ifx \showCODEN    \undefined \def \showCODEN     #1{\unskip}     \fi
\ifx \showDOI      \undefined \def \showDOI       #1{#1}\fi
\ifx \showISBNx    \undefined \def \showISBNx     #1{\unskip}     \fi
\ifx \showISBNxiii \undefined \def \showISBNxiii  #1{\unskip}     \fi
\ifx \showISSN     \undefined \def \showISSN      #1{\unskip}     \fi
\ifx \showLCCN     \undefined \def \showLCCN      #1{\unskip}     \fi
\ifx \shownote     \undefined \def \shownote      #1{#1}          \fi
\ifx \showarticletitle \undefined \def \showarticletitle #1{#1}   \fi
\ifx \showURL      \undefined \def \showURL       {\relax}        \fi
\providecommand\bibfield[2]{#2}
\providecommand\bibinfo[2]{#2}
\providecommand\natexlab[1]{#1}
\providecommand\showeprint[2][]{arXiv:#2}

\bibitem[\protect\citeauthoryear{??}{CWE}{2020}]%
        {CWETop25}
 \bibinfo{year}{2020}\natexlab{}.
\newblock \bibinfo{title}{CWE Top 25 Most Dangerous Software Errors}.
\newblock
  \bibinfo{howpublished}{{https://cwe.mitre.org/top25/archive/2019/2019{\_}cwe{\_}top25.html}}.
\newblock


\bibitem[\protect\citeauthoryear{??}{git}{2020}]%
        {gitlab-survey}
 \bibinfo{year}{2020}\natexlab{}.
\newblock \bibinfo{title}{GitLab: Mapping the DevSecOps Landscape - 2020
  Survey}.
\newblock \bibinfo{howpublished}{{https://about.gitlab.com/developer-survey}}.
\newblock


\bibitem[\protect\citeauthoryear{Ackerman, Buchwald, and Lewski}{Ackerman
  et~al\mbox{.}}{1989}]%
        {Ackerman:1989}
\bibfield{author}{\bibinfo{person}{A. Ackerman}, \bibinfo{person}{L. Buchwald},
  {and} \bibinfo{person}{F. Lewski}.} \bibinfo{year}{1989}\natexlab{}.
\newblock \showarticletitle{Software inspections: an effective verification
  process}.
\newblock \bibinfo{journal}{\emph{IEEE Software}} \bibinfo{volume}{6},
  \bibinfo{number}{3} (\bibinfo{year}{1989}), \bibinfo{pages}{31--36}.
\newblock


\bibitem[\protect\citeauthoryear{Ackerman, Fowler, and Ebenau}{Ackerman
  et~al\mbox{.}}{1984}]%
        {Ackerman:1984}
\bibfield{author}{\bibinfo{person}{A. Ackerman}, \bibinfo{person}{P. Fowler},
  {and} \bibinfo{person}{R. Ebenau}.} \bibinfo{year}{1984}\natexlab{}.
\newblock \showarticletitle{Software Inspections and the Industrial Production
  of Software}. In \bibinfo{booktitle}{\emph{Proceedings of the Symposium on
  Software Validation: Inspection-Testing-Verification-Alternatives}}.
  \bibinfo{pages}{13--40}.
\newblock


\bibitem[\protect\citeauthoryear{Akinola and Osofisan}{Akinola and
  Osofisan}{2009}]%
        {akinola2009empirical}
\bibfield{author}{\bibinfo{person}{O. Akinola} {and} \bibinfo{person}{A.
  Osofisan}.} \bibinfo{year}{2009}\natexlab{}.
\newblock \showarticletitle{An Empirical Comparative Study of Checklist based
  and Ad Hoc Code Reading Techniques in a Distributed Groupware Environment}.
\newblock \bibinfo{journal}{\emph{arXiv preprint arXiv:0909.4260}}
  (\bibinfo{year}{2009}).
\newblock


\bibitem[\protect\citeauthoryear{Alam}{Alam}{2010}]%
        {alam2010software}
\bibfield{author}{\bibinfo{person}{M. Alam}.} \bibinfo{year}{2010}\natexlab{}.
\newblock \showarticletitle{Software security requirements checklist}.
\newblock \bibinfo{journal}{\emph{International Journal of Software
  Engineering}} \bibinfo{volume}{3}, \bibinfo{number}{1}
  (\bibinfo{year}{2010}), \bibinfo{pages}{53--62}.
\newblock


\bibitem[\protect\citeauthoryear{Assal and Chiasson}{Assal and
  Chiasson}{[n.d.]}]%
        {Assal:2018}
\bibfield{author}{\bibinfo{person}{H. Assal} {and} \bibinfo{person}{S.
  Chiasson}.} \bibinfo{year}{[n.d.]}\natexlab{}.
\newblock \showarticletitle{Security in the software development lifecycle}. In
  \bibinfo{booktitle}{\emph{Proceedings of the symposium on usable privacy and
  security}}. \bibinfo{pages}{281--296}.
\newblock


\bibitem[\protect\citeauthoryear{Ayewah and Pugh}{Ayewah and Pugh}{2008}]%
        {Ayewah:2008}
\bibfield{author}{\bibinfo{person}{N. Ayewah} {and} \bibinfo{person}{W. Pugh}.}
  \bibinfo{year}{2008}\natexlab{}.
\newblock \showarticletitle{A report on a survey and study of static analysis
  users}. In \bibinfo{booktitle}{\emph{Proceedings of the workshop on Defects
  in large software systems}}. \bibinfo{pages}{1--5}.
\newblock


\bibitem[\protect\citeauthoryear{Ayewah, Pugh, Hovemeyer, Morgenthaler, and
  Penix}{Ayewah et~al\mbox{.}}{2008}]%
        {Ayewah:2008-2}
\bibfield{author}{\bibinfo{person}{N. Ayewah}, \bibinfo{person}{W. Pugh},
  \bibinfo{person}{D. Hovemeyer}, \bibinfo{person}{J. Morgenthaler}, {and}
  \bibinfo{person}{J. Penix}.} \bibinfo{year}{2008}\natexlab{}.
\newblock \showarticletitle{Using static analysis to find bugs}.
\newblock \bibinfo{journal}{\emph{IEEE software}} \bibinfo{volume}{25},
  \bibinfo{number}{5} (\bibinfo{year}{2008}), \bibinfo{pages}{22--29}.
\newblock


\bibitem[\protect\citeauthoryear{Bacchelli and Bird}{Bacchelli and
  Bird}{2013}]%
        {Bacchelli:2013}
\bibfield{author}{\bibinfo{person}{A. Bacchelli} {and} \bibinfo{person}{C.
  Bird}.} \bibinfo{year}{2013}\natexlab{}.
\newblock \showarticletitle{Expectations, outcomes, and challenges of modern
  code review}. In \bibinfo{booktitle}{\emph{Proceedings of the International
  Conference on Software Engineering}}. \bibinfo{pages}{712--721}.
\newblock


\bibitem[\protect\citeauthoryear{Balfanz, Durfee, Smetters, and
  Grinter}{Balfanz et~al\mbox{.}}{2004}]%
        {Balfanz:2004}
\bibfield{author}{\bibinfo{person}{D. Balfanz}, \bibinfo{person}{G. Durfee},
  \bibinfo{person}{D. Smetters}, {and} \bibinfo{person}{R. Grinter}.}
  \bibinfo{year}{2004}\natexlab{}.
\newblock \showarticletitle{In search of usable security: Five lessons from the
  field}.
\newblock \bibinfo{journal}{\emph{IEEE Security \& Privacy}}
  \bibinfo{volume}{2}, \bibinfo{number}{5} (\bibinfo{year}{2004}),
  \bibinfo{pages}{19--24}.
\newblock


\bibitem[\protect\citeauthoryear{Baum, Liskin, Niklas, and Schneider}{Baum
  et~al\mbox{.}}{2016}]%
        {baum2016factors}
\bibfield{author}{\bibinfo{person}{T. Baum}, \bibinfo{person}{O. Liskin},
  \bibinfo{person}{K. Niklas}, {and} \bibinfo{person}{K. Schneider}.}
  \bibinfo{year}{2016}\natexlab{}.
\newblock \showarticletitle{Factors influencing code review processes in
  industry}. In \bibinfo{booktitle}{\emph{Proceedings of the international
  symposium on foundations of software engineering}}. \bibinfo{pages}{85--96}.
\newblock


\bibitem[\protect\citeauthoryear{Baum, Schneider, and Bacchelli}{Baum
  et~al\mbox{.}}{2019}]%
        {Baum:2019}
\bibfield{author}{\bibinfo{person}{T. Baum}, \bibinfo{person}{K. Schneider},
  {and} \bibinfo{person}{A. Bacchelli}.} \bibinfo{year}{2019}\natexlab{}.
\newblock \showarticletitle{Associating working memory capacity and code change
  ordering with code review performance}.
\newblock \bibinfo{journal}{\emph{Empirical Software Engineering}}
  \bibinfo{volume}{24}, \bibinfo{number}{4} (\bibinfo{year}{2019}),
  \bibinfo{pages}{1762--1798}.
\newblock


\bibitem[\protect\citeauthoryear{Bellovin}{Bellovin}{2008}]%
        {Bellovin:2008}
\bibfield{author}{\bibinfo{person}{S. Bellovin}.}
  \bibinfo{year}{2008}\natexlab{}.
\newblock \showarticletitle{Security by Checklist}.
\newblock \bibinfo{journal}{\emph{IEEE Security Privacy}} \bibinfo{volume}{6},
  \bibinfo{number}{2} (\bibinfo{year}{2008}), \bibinfo{pages}{88--88}.
\newblock


\bibitem[\protect\citeauthoryear{Boehm and Basili}{Boehm and Basili}{2001}]%
        {Boehm:2001}
\bibfield{author}{\bibinfo{person}{B. Boehm} {and} \bibinfo{person}{V.
  Basili}.} \bibinfo{year}{2001}\natexlab{}.
\newblock \showarticletitle{Software Defect Reduction Top 10 List}.
\newblock  \bibinfo{volume}{34}, \bibinfo{number}{1} (\bibinfo{year}{2001}),
  \bibinfo{pages}{135--137}.
\newblock


\bibitem[\protect\citeauthoryear{Braz, {\c{C}}alikli, and Bacchelli}{Braz
  et~al\mbox{.}}{2022}]%
        {replication-package}
\bibfield{author}{\bibinfo{person}{L. Braz}, \bibinfo{person}{G.
  {\c{C}}alikli}, {and} \bibinfo{person}{A. Bacchelli}.}
  \bibinfo{year}{2022}\natexlab{}.
\newblock \bibinfo{title}{Data and Material}.
\newblock \bibinfo{howpublished}{\url{https://doi.org/10.5281/zenodo.6026291}}.
\newblock


\bibitem[\protect\citeauthoryear{Braz, Fregnan, {\c{C}}alikli, and
  Bacchelli}{Braz et~al\mbox{.}}{2021}]%
        {braz:2021}
\bibfield{author}{\bibinfo{person}{L. Braz}, \bibinfo{person}{E. Fregnan},
  \bibinfo{person}{G. {\c{C}}alikli}, {and} \bibinfo{person}{A. Bacchelli}.}
  \bibinfo{year}{2021}\natexlab{}.
\newblock \showarticletitle{Why Don't Developers Detect Improper Input
  Validation?'; DROP TABLE Papers;--}. In \bibinfo{booktitle}{\emph{Proceedings
  of the International Conference on Software Engineering}}.
  \bibinfo{pages}{499--511}.
\newblock


\bibitem[\protect\citeauthoryear{Brykczynski}{Brykczynski}{1999}]%
        {brykczynski1999survey}
\bibfield{author}{\bibinfo{person}{B. Brykczynski}.}
  \bibinfo{year}{1999}\natexlab{}.
\newblock \showarticletitle{A survey of software inspection checklists}.
\newblock \bibinfo{journal}{\emph{ACM SIGSOFT Software Engineering Notes}}
  \bibinfo{volume}{24}, \bibinfo{number}{1} (\bibinfo{year}{1999}),
  \bibinfo{pages}{82}.
\newblock


\bibitem[\protect\citeauthoryear{Chernak}{Chernak}{1996}]%
        {chernak1996statistical}
\bibfield{author}{\bibinfo{person}{Y. Chernak}.}
  \bibinfo{year}{1996}\natexlab{}.
\newblock \showarticletitle{A statistical approach to the inspection checklist
  formal synthesis and improvement}.
\newblock \bibinfo{journal}{\emph{Transactions on Software Engineering}}
  \bibinfo{volume}{22}, \bibinfo{number}{12} (\bibinfo{year}{1996}),
  \bibinfo{pages}{866--874}.
\newblock


\bibitem[\protect\citeauthoryear{Chong, Thongtanunam, and
  Tantithamthavorn}{Chong et~al\mbox{.}}{2021}]%
        {chong2021assessing}
\bibfield{author}{\bibinfo{person}{C. Chong}, \bibinfo{person}{P.
  Thongtanunam}, {and} \bibinfo{person}{C. Tantithamthavorn}.}
  \bibinfo{year}{2021}\natexlab{}.
\newblock \showarticletitle{Assessing the Students' Understanding and their
  Mistakes in Code Review Checklists-An Experience Report of 1,791 Code Review
  Checklist Questions from 394 Students}. In
  \bibinfo{booktitle}{\emph{Proceedings of the International Conference on
  Software Engineering: Software Engineering Education and Training}}.
  \bibinfo{pages}{20--29}.
\newblock


\bibitem[\protect\citeauthoryear{Christakis and Bird}{Christakis and
  Bird}{2016}]%
        {Christakis:2016}
\bibfield{author}{\bibinfo{person}{M. Christakis} {and} \bibinfo{person}{C.
  Bird}.} \bibinfo{year}{2016}\natexlab{}.
\newblock \showarticletitle{What developers want and need from program
  analysis: an empirical study}. In \bibinfo{booktitle}{\emph{Proceedings of
  the international conference on automated software engineering}}.
  \bibinfo{pages}{332--343}.
\newblock


\bibitem[\protect\citeauthoryear{Cohen}{Cohen}{2010}]%
        {Cohe2010a}
\bibfield{author}{\bibinfo{person}{J. Cohen}.} \bibinfo{year}{2010}\natexlab{}.
\newblock \showarticletitle{Modern Code Review}.
\newblock In \bibinfo{booktitle}{\emph{Making Software}}.
  \bibinfo{publisher}{O'Reilly}, Chapter~18, \bibinfo{pages}{329--338}.
\newblock


\bibitem[\protect\citeauthoryear{{Cook} and {Campbell}}{{Cook} and
  {Campbell}}{1979}]%
        {Cook:1979}
\bibfield{author}{\bibinfo{person}{T. {Cook}} {and} \bibinfo{person}{D.
  {Campbell}}.} \bibinfo{year}{1979}\natexlab{}.
\newblock \bibinfo{booktitle}{\emph{Quasi-Experimentation: Design and Analysis
  Issues for Field Settings}}.
\newblock \bibinfo{publisher}{Houghton Mifflin Company}.
\newblock


\bibitem[\protect\citeauthoryear{Cook, Campbell, and Shadish}{Cook
  et~al\mbox{.}}{2002}]%
        {cook2002experimental}
\bibfield{author}{\bibinfo{person}{T. Cook}, \bibinfo{person}{D. Campbell},
  {and} \bibinfo{person}{W. Shadish}.} \bibinfo{year}{2002}\natexlab{}.
\newblock \bibinfo{booktitle}{\emph{Experimental and quasi-experimental designs
  for generalized causal inference}}.
\newblock


\bibitem[\protect\citeauthoryear{Cruzes, Felderer, Oyetoyan, Gander, and
  Pekaric}{Cruzes et~al\mbox{.}}{2017}]%
        {Cruzes:2017}
\bibfield{author}{\bibinfo{person}{D. Cruzes}, \bibinfo{person}{M. Felderer},
  \bibinfo{person}{T. Oyetoyan}, \bibinfo{person}{M. Gander}, {and}
  \bibinfo{person}{I. Pekaric}.} \bibinfo{year}{2017}\natexlab{}.
\newblock \showarticletitle{How is security testing done in agile teams? A
  cross-case analysis of four software teams}. In
  \bibinfo{booktitle}{\emph{Proceedings of the International Conference on
  Agile Software Development}}. Springer, Cham, \bibinfo{pages}{201--216}.
\newblock


\bibitem[\protect\citeauthoryear{Dunsmore, Roper, and Wood}{Dunsmore
  et~al\mbox{.}}{2003a}]%
        {dunsmore2003development}
\bibfield{author}{\bibinfo{person}{A. Dunsmore}, \bibinfo{person}{M. Roper},
  {and} \bibinfo{person}{M. Wood}.} \bibinfo{year}{2003}\natexlab{a}.
\newblock \showarticletitle{The development and evaluation of three diverse
  techniques for object-oriented code inspection}.
\newblock \bibinfo{journal}{\emph{Transactions on Software Engineering}}
  \bibinfo{volume}{29}, \bibinfo{number}{8} (\bibinfo{year}{2003}),
  \bibinfo{pages}{677--686}.
\newblock


\bibitem[\protect\citeauthoryear{Dunsmore, Roper, and Wood}{Dunsmore
  et~al\mbox{.}}{2003b}]%
        {Dunsmore:2003}
\bibfield{author}{\bibinfo{person}{A. Dunsmore}, \bibinfo{person}{M. Roper},
  {and} \bibinfo{person}{M. Wood}.} \bibinfo{year}{2003}\natexlab{b}.
\newblock \showarticletitle{The Development and Evaluation of Three Diverse
  Techniques for Object-Oriented Code Inspection}.
\newblock \bibinfo{journal}{\emph{Transactions on Software Engineering}}
  \bibinfo{volume}{29}, \bibinfo{number}{8} (\bibinfo{year}{2003}),
  \bibinfo{pages}{677–686}.
\newblock


\bibitem[\protect\citeauthoryear{Edmundson, Holtkamp, Rivera, Finifter,
  Mettler, and Wagner}{Edmundson et~al\mbox{.}}{2013}]%
        {edmundson:2013}
\bibfield{author}{\bibinfo{person}{A. Edmundson}, \bibinfo{person}{B.
  Holtkamp}, \bibinfo{person}{E. Rivera}, \bibinfo{person}{M. Finifter},
  \bibinfo{person}{A. Mettler}, {and} \bibinfo{person}{D. Wagner}.}
  \bibinfo{year}{2013}\natexlab{}.
\newblock \showarticletitle{An empirical study on the effectiveness of security
  code review}. In \bibinfo{booktitle}{\emph{Proceedings of the International
  Symposium on Engineering Secure Software and Systems}}.
  \bibinfo{pages}{197--212}.
\newblock


\bibitem[\protect\citeauthoryear{Fagan}{Fagan}{1976}]%
        {fagan1976design}
\bibfield{author}{\bibinfo{person}{M. Fagan}.} \bibinfo{year}{1976}\natexlab{}.
\newblock \showarticletitle{Design and code inspections to reduce errors in
  program development}.
\newblock \bibinfo{journal}{\emph{IBM Systems Journal}} \bibinfo{volume}{15},
  \bibinfo{number}{3} (\bibinfo{year}{1976}), \bibinfo{pages}{182--211}.
\newblock


\bibitem[\protect\citeauthoryear{Fagan}{Fagan}{2002}]%
        {fagan2002design}
\bibfield{author}{\bibinfo{person}{M. Fagan}.} \bibinfo{year}{2002}\natexlab{}.
\newblock \showarticletitle{Design and code inspections to reduce errors in
  program development}.
\newblock In \bibinfo{booktitle}{\emph{Software pioneers}}.
  \bibinfo{publisher}{Springer}, \bibinfo{pages}{575--607}.
\newblock


\bibitem[\protect\citeauthoryear{Falessi, Juristo, Wohlin, Turhan, M{\"{u}}nch,
  Jedlitschka, and Oivo}{Falessi et~al\mbox{.}}{2018}]%
        {Falessi:2018}
\bibfield{author}{\bibinfo{person}{D. Falessi}, \bibinfo{person}{N. Juristo},
  \bibinfo{person}{C. Wohlin}, \bibinfo{person}{B. Turhan}, \bibinfo{person}{J.
  M{\"{u}}nch}, \bibinfo{person}{A. Jedlitschka}, {and} \bibinfo{person}{M.
  Oivo}.} \bibinfo{year}{2018}\natexlab{}.
\newblock \showarticletitle{Empirical Software Engineering Experts on the Use
  of Students and Professionals in Experiments}.
\newblock \bibinfo{journal}{\emph{Empirical Software Engineering}}
  \bibinfo{volume}{23}, \bibinfo{number}{1} (\bibinfo{year}{2018}),
  \bibinfo{pages}{452--489}.
\newblock


\bibitem[\protect\citeauthoryear{Faul, Erfelder, Lang, and Buchner}{Faul
  et~al\mbox{.}}{2007}]%
        {Faul:2007}
\bibfield{author}{\bibinfo{person}{F. Faul}, \bibinfo{person}{E. Erfelder},
  \bibinfo{person}{A.~G. Lang}, {and} \bibinfo{person}{A. Buchner}.}
  \bibinfo{year}{2007}\natexlab{}.
\newblock \showarticletitle{GPower3: A flexible statistical power analysis
  program for the social, behavioral, and biomedical sciences}.
\newblock \bibinfo{journal}{\emph{Behavior Research Methods}}
  \bibinfo{volume}{39} (\bibinfo{year}{2007}), \bibinfo{pages}{175 -- 191}.
\newblock


\bibitem[\protect\citeauthoryear{Foundation}{Foundation}{2017}]%
        {owasp}
\bibfield{author}{\bibinfo{person}{The~OWASP Foundation}.}
  \bibinfo{year}{2017}\natexlab{}.
\newblock \bibinfo{booktitle}{\emph{OWASP Foundation}}.
\newblock
\urldef\tempurl%
\url{https://owasp.org/}
\showURL{%
Retrieved July 18, 2021 from \tempurl}


\bibitem[\protect\citeauthoryear{Garrison and Posey}{Garrison and
  Posey}{2006}]%
        {garrison2006computer}
\bibfield{author}{\bibinfo{person}{C. Garrison} {and} \bibinfo{person}{R.
  Posey}.} \bibinfo{year}{2006}\natexlab{}.
\newblock \showarticletitle{Computer security checklist for non-security
  technology professionals}.
\newblock \bibinfo{journal}{\emph{Journal of International Technology and
  Information Management}} \bibinfo{volume}{15}, \bibinfo{number}{3}
  (\bibinfo{year}{2006}), \bibinfo{pages}{7}.
\newblock


\bibitem[\protect\citeauthoryear{Gerrit}{Gerrit}{2020}]%
        {Gerrit}
\bibfield{author}{\bibinfo{person}{Gerrit}.} \bibinfo{year}{2020}\natexlab{}.
\newblock \bibinfo{title}{Gerrit Code Review}.
\newblock \bibinfo{howpublished}{{https://www.gerritcodereview.com/}}.
\newblock


\bibitem[\protect\citeauthoryear{Gilb and Graham}{Gilb and Graham}{1993}]%
        {gilb1993software}
\bibfield{author}{\bibinfo{person}{T. Gilb} {and} \bibinfo{person}{D. Graham}.}
  \bibinfo{year}{1993}\natexlab{}.
\newblock \bibinfo{booktitle}{\emph{Software inspections}}.
\newblock \bibinfo{publisher}{Addison-Wesley Reading, Masachusetts}.
\newblock


\bibitem[\protect\citeauthoryear{Gilliam, Wolfe, Sherif, and Bishop}{Gilliam
  et~al\mbox{.}}{2003}]%
        {gilliam2003software}
\bibfield{author}{\bibinfo{person}{D. Gilliam}, \bibinfo{person}{T. Wolfe},
  \bibinfo{person}{J. Sherif}, {and} \bibinfo{person}{M. Bishop}.}
  \bibinfo{year}{2003}\natexlab{}.
\newblock \showarticletitle{Software security checklist for the software life
  cycle}. In \bibinfo{booktitle}{\emph{Proceedings of the International
  Workshops on Enabling Technologies: Infrastructure for Collaborative
  Enterprises}}. \bibinfo{pages}{243--248}.
\newblock


\bibitem[\protect\citeauthoryear{Gon{\c{c}}alves, Fregnan, Baum, Schneider, and
  Bacchelli}{Gon{\c{c}}alves et~al\mbox{.}}{2020}]%
        {gonccalves2020explicit}
\bibfield{author}{\bibinfo{person}{P. Gon{\c{c}}alves}, \bibinfo{person}{E.
  Fregnan}, \bibinfo{person}{T. Baum}, \bibinfo{person}{K. Schneider}, {and}
  \bibinfo{person}{A. Bacchelli}.} \bibinfo{year}{2020}\natexlab{}.
\newblock \showarticletitle{Do Explicit Review Strategies Improve Code Review
  Performance?}. In \bibinfo{booktitle}{\emph{Proceedings of the International
  Conference on Mining Software Repositories}}. \bibinfo{pages}{606--610}.
\newblock


\bibitem[\protect\citeauthoryear{Green and Smith}{Green and Smith}{2016}]%
        {Green:2016}
\bibfield{author}{\bibinfo{person}{M. Green} {and} \bibinfo{person}{M. Smith}.}
  \bibinfo{year}{2016}\natexlab{}.
\newblock \showarticletitle{Developers are not the enemy!: The need for usable
  security apis}.
\newblock \bibinfo{journal}{\emph{IEEE Security \& Privacy}}
  \bibinfo{volume}{14}, \bibinfo{number}{5} (\bibinfo{year}{2016}),
  \bibinfo{pages}{40--46}.
\newblock


\bibitem[\protect\citeauthoryear{HackerOne}{HackerOne}{2022}]%
        {HackerOne}
\bibfield{author}{\bibinfo{person}{HackerOne}.}
  \bibinfo{year}{2022}\natexlab{}.
\newblock \bibinfo{title}{HackerOne}.
\newblock \bibinfo{howpublished}{https://www.hackerone.com/}.
\newblock


\bibitem[\protect\citeauthoryear{Hermans}{Hermans}{2021}]%
        {hermans2021programmer}
\bibfield{author}{\bibinfo{person}{F. Hermans}.}
  \bibinfo{year}{2021}\natexlab{}.
\newblock \bibinfo{booktitle}{\emph{The Programmer's Brain: What every
  programmer needs to know about cognition}}.
\newblock \bibinfo{publisher}{Manning Publications}.
\newblock


\bibitem[\protect\citeauthoryear{Kollanus and Koskinen}{Kollanus and
  Koskinen}{2009}]%
        {kollanus2009survey}
\bibfield{author}{\bibinfo{person}{S. Kollanus} {and} \bibinfo{person}{J.
  Koskinen}.} \bibinfo{year}{2009}\natexlab{}.
\newblock \showarticletitle{Survey of software inspection research}.
\newblock \bibinfo{journal}{\emph{The Open Software Engineering Journal}}
  \bibinfo{volume}{3}, \bibinfo{number}{1} (\bibinfo{year}{2009}).
\newblock


\bibitem[\protect\citeauthoryear{Laitenberger and DeBaud}{Laitenberger and
  DeBaud}{2000}]%
        {laitenberger:2000}
\bibfield{author}{\bibinfo{person}{O. Laitenberger} {and} \bibinfo{person}{J.
  DeBaud}.} \bibinfo{year}{2000}\natexlab{}.
\newblock \showarticletitle{An encompassing life cycle centric survey of
  software inspection}.
\newblock \bibinfo{journal}{\emph{Journal of systems and software}}
  \bibinfo{volume}{50}, \bibinfo{number}{1} (\bibinfo{year}{2000}),
  \bibinfo{pages}{5--31}.
\newblock


\bibitem[\protect\citeauthoryear{Lanubile and Visaggio}{Lanubile and
  Visaggio}{2000}]%
        {lanubile2000evaluating}
\bibfield{author}{\bibinfo{person}{F. Lanubile} {and} \bibinfo{person}{G.
  Visaggio}.} \bibinfo{year}{2000}\natexlab{}.
\newblock \showarticletitle{Evaluating defect detection techniques for software
  requirements inspections}.
\newblock \bibinfo{journal}{\emph{International SoftwareEngineering Research
  Network, Report}} (\bibinfo{year}{2000}), \bibinfo{pages}{1--24}.
\newblock


\bibitem[\protect\citeauthoryear{M{\"a}ntyl{\"a} and Lassenius}{M{\"a}ntyl{\"a}
  and Lassenius}{2008}]%
        {mantyla2008types}
\bibfield{author}{\bibinfo{person}{M. M{\"a}ntyl{\"a}} {and}
  \bibinfo{person}{C. Lassenius}.} \bibinfo{year}{2008}\natexlab{}.
\newblock \showarticletitle{What types of defects are really discovered in code
  reviews?}
\newblock \bibinfo{journal}{\emph{Transactions on Software Engineering}}
  \bibinfo{volume}{35}, \bibinfo{number}{3} (\bibinfo{year}{2008}),
  \bibinfo{pages}{430--448}.
\newblock


\bibitem[\protect\citeauthoryear{McGraw}{McGraw}{2004}]%
        {McGraw:2004}
\bibfield{author}{\bibinfo{person}{G. McGraw}.}
  \bibinfo{year}{2004}\natexlab{}.
\newblock \showarticletitle{Software security}.
\newblock \bibinfo{journal}{\emph{IEEE Security Privacy}} \bibinfo{volume}{2},
  \bibinfo{number}{2} (\bibinfo{year}{2004}), \bibinfo{pages}{80--83}.
\newblock


\bibitem[\protect\citeauthoryear{{Mcintosh}, {Kamei}, {Adams}, and
  {Hassan}}{{Mcintosh} et~al\mbox{.}}{2016}]%
        {Mcintosh:2016}
\bibfield{author}{\bibinfo{person}{S. {Mcintosh}}, \bibinfo{person}{Y.
  {Kamei}}, \bibinfo{person}{B. {Adams}}, {and} \bibinfo{person}{A.~E.
  {Hassan}}.} \bibinfo{year}{2016}\natexlab{}.
\newblock \showarticletitle{An Empirical Study of the Impact of Modern Code
  Review Practices on Software Quality}.
\newblock \bibinfo{journal}{\emph{Empirical Software Engineering}}
  \bibinfo{volume}{21}, \bibinfo{number}{5} (\bibinfo{year}{2016}),
  \bibinfo{pages}{2146--2189}.
\newblock


\bibitem[\protect\citeauthoryear{Meneely and Williams}{Meneely and
  Williams}{2010}]%
        {Meneely:2010}
\bibfield{author}{\bibinfo{person}{A. Meneely} {and} \bibinfo{person}{L.
  Williams}.} \bibinfo{year}{2010}\natexlab{}.
\newblock \showarticletitle{Strengthening the Empirical Analysis of the
  Relationship between Linus' Law and Software Security}. In
  \bibinfo{booktitle}{\emph{Proceedings of the International Symposium on
  Empirical Software Engineering and Measurement}}. \bibinfo{pages}{1--10}.
\newblock


\bibitem[\protect\citeauthoryear{Meneely and Williams}{Meneely and
  Williams}{2012}]%
        {Meneely:2012}
\bibfield{author}{\bibinfo{person}{A. Meneely} {and} \bibinfo{person}{O.
  Williams}.} \bibinfo{year}{2012}\natexlab{}.
\newblock \showarticletitle{Interactive Churn Metrics: Socio-Technical Variants
  of Code Churn}.
\newblock \bibinfo{journal}{\emph{Software Engineering Notes}}
  \bibinfo{volume}{37}, \bibinfo{number}{6} (\bibinfo{year}{2012}),
  \bibinfo{pages}{1--6}.
\newblock


\bibitem[\protect\citeauthoryear{Naiakshina, Danilova, Gerlitz, von Zezschwitz,
  and Smith}{Naiakshina et~al\mbox{.}}{2019}]%
        {Naiakshina:2019}
\bibfield{author}{\bibinfo{person}{A. Naiakshina}, \bibinfo{person}{A.
  Danilova}, \bibinfo{person}{E. Gerlitz}, \bibinfo{person}{E. von Zezschwitz},
  {and} \bibinfo{person}{M. Smith}.} \bibinfo{year}{2019}\natexlab{}.
\newblock \bibinfo{booktitle}{\emph{"If You Want, I Can Store the Encrypted
  Password": A Password-Storage Field Study with Freelance Developers}}.
\newblock \bibinfo{pages}{1--12}.
\newblock


\bibitem[\protect\citeauthoryear{Naiakshina, Danilova, Tiefenau, Herzog,
  Dechand, and Smith}{Naiakshina et~al\mbox{.}}{2017}]%
        {Naiakshina:2017}
\bibfield{author}{\bibinfo{person}{A. Naiakshina}, \bibinfo{person}{A.
  Danilova}, \bibinfo{person}{C. Tiefenau}, \bibinfo{person}{M. Herzog},
  \bibinfo{person}{S. Dechand}, {and} \bibinfo{person}{M. Smith}.}
  \bibinfo{year}{2017}\natexlab{}.
\newblock \showarticletitle{Why Do Developers Get Password Storage Wrong? A
  Qualitative Usability Study}. In \bibinfo{booktitle}{\emph{Proceedings of the
  Conference on Computer and Communications Security}}.
  \bibinfo{pages}{311--328}.
\newblock


\bibitem[\protect\citeauthoryear{Naiakshina, Danilova, Tiefenau, and
  Smith}{Naiakshina et~al\mbox{.}}{2018}]%
        {Naiakshina:2018}
\bibfield{author}{\bibinfo{person}{A. Naiakshina}, \bibinfo{person}{A.
  Danilova}, \bibinfo{person}{C. Tiefenau}, {and} \bibinfo{person}{M. Smith}.}
  \bibinfo{year}{2018}\natexlab{}.
\newblock \showarticletitle{Deception Task Design in Developer Password
  Studies: Exploring a Student Sample}. In
  \bibinfo{booktitle}{\emph{Proceedings of the Conference on Usable Privacy and
  Security}}. \bibinfo{pages}{297--313}.
\newblock


\bibitem[\protect\citeauthoryear{Neumayer and Pl{\"u}mper}{Neumayer and
  Pl{\"u}mper}{2017}]%
        {neumayer:2017}
\bibfield{author}{\bibinfo{person}{E. Neumayer} {and} \bibinfo{person}{T.
  Pl{\"u}mper}.} \bibinfo{year}{2017}\natexlab{}.
\newblock \bibinfo{booktitle}{\emph{Robustness tests for quantitative
  research}}.
\newblock \bibinfo{publisher}{Cambridge University Press}.
\newblock


\bibitem[\protect\citeauthoryear{Oladele and Adedayo}{Oladele and
  Adedayo}{2014}]%
        {oladele2014empirical}
\bibfield{author}{\bibinfo{person}{R. Oladele} {and} \bibinfo{person}{H.
  Adedayo}.} \bibinfo{year}{2014}\natexlab{}.
\newblock \showarticletitle{On empirical comparison of checklist-based reading
  and adhoc reading for code inspection}.
\newblock \bibinfo{journal}{\emph{International Journal of Computer
  Applications}} \bibinfo{volume}{87}, \bibinfo{number}{1}
  (\bibinfo{year}{2014}).
\newblock


\bibitem[\protect\citeauthoryear{Pieczul, Foley, and Zurko}{Pieczul
  et~al\mbox{.}}{2017}]%
        {Pieczul:2017}
\bibfield{author}{\bibinfo{person}{O. Pieczul}, \bibinfo{person}{S. Foley},
  {and} \bibinfo{person}{M. Zurko}.} \bibinfo{year}{2017}\natexlab{}.
\newblock \showarticletitle{Developer-Centered Security and the Symmetry of
  Ignorance}. In \bibinfo{booktitle}{\emph{Proceedings of the New Security
  Paradigms Workshop}}. \bibinfo{publisher}{Association for Computing
  Machinery}, \bibinfo{pages}{46–56}.
\newblock


\bibitem[\protect\citeauthoryear{Planning}{Planning}{2002}]%
        {planning2002economic}
\bibfield{author}{\bibinfo{person}{Strategic Planning}.}
  \bibinfo{year}{2002}\natexlab{}.
\newblock \showarticletitle{The economic impacts of inadequate infrastructure
  for software testing}.
\newblock \bibinfo{journal}{\emph{National Institute of Standards and
  Technology}} (\bibinfo{year}{2002}).
\newblock


\bibitem[\protect\citeauthoryear{Poller, Kocksch, Trpe, Epp, and
  Kinder-Kurlanda}{Poller et~al\mbox{.}}{2017}]%
        {Poller:2017}
\bibfield{author}{\bibinfo{person}{A. Poller}, \bibinfo{person}{L. Kocksch},
  \bibinfo{person}{S. Trpe}, \bibinfo{person}{F. Epp}, {and}
  \bibinfo{person}{K. Kinder-Kurlanda}.} \bibinfo{year}{2017}\natexlab{}.
\newblock \showarticletitle{Can security become a routine? A study of
  organizational change in an agile software development group}. In
  \bibinfo{booktitle}{\emph{Proceedings of the conference on computer supported
  cooperative work and social computing}}. \bibinfo{pages}{2489--2503}.
\newblock


\bibitem[\protect\citeauthoryear{Porter, Votta, and Basili}{Porter
  et~al\mbox{.}}{1995}]%
        {porter1995comparing}
\bibfield{author}{\bibinfo{person}{A. Porter}, \bibinfo{person}{L.~G Votta},
  {and} \bibinfo{person}{V. Basili}.} \bibinfo{year}{1995}\natexlab{}.
\newblock \showarticletitle{Comparing detection methods for software
  requirements inspections: A replicated experiment}.
\newblock \bibinfo{journal}{\emph{Transactions on Software Engineering}}
  \bibinfo{volume}{21}, \bibinfo{number}{6} (\bibinfo{year}{1995}),
  \bibinfo{pages}{563--575}.
\newblock


\bibitem[\protect\citeauthoryear{Project}{Project}{2013}]%
        {CWESDE}
\bibfield{author}{\bibinfo{person}{CWE Project}.}
  \bibinfo{year}{2013}\natexlab{}.
\newblock \bibinfo{booktitle}{\emph{CWE Category - Sensitive Data Exposure}}.
\newblock
\urldef\tempurl%
\url{https://cwe.mitre.org/data/definitions/1029.html}
\showURL{%
Retrieved July 19, 2021 from \tempurl}


\bibitem[\protect\citeauthoryear{Project}{Project}{2021a}]%
        {CWELogExample}
\bibfield{author}{\bibinfo{person}{CWE Project}.}
  \bibinfo{year}{2021}\natexlab{a}.
\newblock \bibinfo{booktitle}{\emph{CWE-209: Generation of Error Message
  Containing Sensitive Information}}.
\newblock
\urldef\tempurl%
\url{https://cwe.mitre.org/data/definitions/209.html}
\showURL{%
Retrieved June 28, 2021 from \tempurl}


\bibitem[\protect\citeauthoryear{Project}{Project}{2021b}]%
        {missingEncrypt}
\bibfield{author}{\bibinfo{person}{CWE Project}.}
  \bibinfo{year}{2021}\natexlab{b}.
\newblock \bibinfo{booktitle}{\emph{CWE-311: Missing Encryption of Sensitive
  Data}}.
\newblock
\urldef\tempurl%
\url{https://cwe.mitre.org/data/definitions/311.html}
\showURL{%
Retrieved August, 2021 from \tempurl}


\bibitem[\protect\citeauthoryear{Project}{Project}{2021c}]%
        {CWEBrokenCrypto}
\bibfield{author}{\bibinfo{person}{CWE Project}.}
  \bibinfo{year}{2021}\natexlab{c}.
\newblock \bibinfo{booktitle}{\emph{CWE-327: Use of a Broken or Risky
  Cryptographic Algorithm}}.
\newblock
\urldef\tempurl%
\url{https://cwe.mitre.org/data/definitions/327.html}
\showURL{%
Retrieved June 29, 2021 from \tempurl}


\bibitem[\protect\citeauthoryear{Project}{Project}{2017a}]%
        {owaspChecklist}
\bibfield{author}{\bibinfo{person}{OWASP Project}.}
  \bibinfo{year}{2017}\natexlab{a}.
\newblock \bibinfo{booktitle}{\emph{OWASP Code Review Guide 2.0}}.
\newblock
\urldef\tempurl%
\url{https://owasp.org/www-pdf-archive/OWASP_Code_Review_Guide_v2.pdf}
\showURL{%
Retrieved June 28, 2021 from \tempurl}


\bibitem[\protect\citeauthoryear{Project}{Project}{2017b}]%
        {owaspTop10}
\bibfield{author}{\bibinfo{person}{OWASP Project}.}
  \bibinfo{year}{2017}\natexlab{b}.
\newblock \bibinfo{booktitle}{\emph{OWASP Top Ten}}.
\newblock
\urldef\tempurl%
\url{https://owasp.org/www-project-top-ten}
\showURL{%
Retrieved May 27, 2021 from \tempurl}


\bibitem[\protect\citeauthoryear{Project}{Project}{2017c}]%
        {sensitiveExposure}
\bibfield{author}{\bibinfo{person}{OWASP Project}.}
  \bibinfo{year}{2017}\natexlab{c}.
\newblock \bibinfo{booktitle}{\emph{Sensitive Data Exposure}}.
\newblock
\urldef\tempurl%
\url{https://owasp.org/www-project-top-ten/2017/A3_2017-Sensitive_Data_Exposure}
\showURL{%
Retrieved June 28, 2021 from \tempurl}


\bibitem[\protect\citeauthoryear{Rigby and Bird}{Rigby and Bird}{2013}]%
        {rigby2013convergent}
\bibfield{author}{\bibinfo{person}{P. Rigby} {and} \bibinfo{person}{C. Bird}.}
  \bibinfo{year}{2013}\natexlab{}.
\newblock \showarticletitle{Convergent contemporary software peer review
  practices}. In \bibinfo{booktitle}{\emph{Proceedings of the Joint Meeting on
  Foundations of Software Engineering}}. \bibinfo{pages}{202--212}.
\newblock


\bibitem[\protect\citeauthoryear{Rigby, German, Cowen, and Storey}{Rigby
  et~al\mbox{.}}{2014}]%
        {rigby2014peer}
\bibfield{author}{\bibinfo{person}{P. Rigby}, \bibinfo{person}{D. German},
  \bibinfo{person}{L. Cowen}, {and} \bibinfo{person}{M. Storey}.}
  \bibinfo{year}{2014}\natexlab{}.
\newblock \showarticletitle{Peer review on open-source software projects:
  Parameters, statistical models, and theory}.
\newblock \bibinfo{journal}{\emph{Transactions on Software Engineering and
  Methodology}} \bibinfo{volume}{23}, \bibinfo{number}{4}
  (\bibinfo{year}{2014}), \bibinfo{pages}{1--33}.
\newblock


\bibitem[\protect\citeauthoryear{Rong, Li, Xie, and Zheng}{Rong
  et~al\mbox{.}}{2012}]%
        {Rong2012}
\bibfield{author}{\bibinfo{person}{G. Rong}, \bibinfo{person}{J. Li},
  \bibinfo{person}{M. Xie}, {and} \bibinfo{person}{T. Zheng}.}
  \bibinfo{year}{2012}\natexlab{}.
\newblock \showarticletitle{The effect of checklist in code review for
  inexperienced students: An empirical study}. In
  \bibinfo{booktitle}{\emph{Proceedings of the Conference on Software
  Engineering Education and Training}}. \bibinfo{pages}{120--124}.
\newblock


\bibitem[\protect\citeauthoryear{Sadowski, S{\"o}derberg, Church, Sipko, and
  Bacchelli}{Sadowski et~al\mbox{.}}{2018}]%
        {sadowski2018modern}
\bibfield{author}{\bibinfo{person}{C. Sadowski}, \bibinfo{person}{E.
  S{\"o}derberg}, \bibinfo{person}{L. Church}, \bibinfo{person}{M. Sipko},
  {and} \bibinfo{person}{A. Bacchelli}.} \bibinfo{year}{2018}\natexlab{}.
\newblock \showarticletitle{Modern code review: a case study at google}. In
  \bibinfo{booktitle}{\emph{Proceedings of the International Conference on
  Software Engineering: Software Engineering in Practice}}.
  \bibinfo{pages}{181--190}.
\newblock


\bibitem[\protect\citeauthoryear{Shin, Meneely, Williams, and Osborne}{Shin
  et~al\mbox{.}}{2011}]%
        {Shin:2011}
\bibfield{author}{\bibinfo{person}{Y. Shin}, \bibinfo{person}{A. Meneely},
  \bibinfo{person}{L. Williams}, {and} \bibinfo{person}{J. Osborne}.}
  \bibinfo{year}{2011}\natexlab{}.
\newblock \showarticletitle{Evaluating Complexity, Code Churn, and Developer
  Activity Metrics as Indicators of Software Vulnerabilities}.
\newblock \bibinfo{journal}{\emph{Transactions on Software Engineering}}
  \bibinfo{volume}{37} (\bibinfo{year}{2011}), \bibinfo{pages}{772--787}.
\newblock


\bibitem[\protect\citeauthoryear{Shirey}{Shirey}{2000}]%
        {shirey2000internet}
\bibfield{author}{\bibinfo{person}{R. Shirey}.}
  \bibinfo{year}{2000}\natexlab{}.
\newblock \bibinfo{title}{Internet security glossary}.
\newblock
\newblock


\bibitem[\protect\citeauthoryear{Shu, Yao, and Bertino}{Shu
  et~al\mbox{.}}{2015}]%
        {shu2015privacy}
\bibfield{author}{\bibinfo{person}{X. Shu}, \bibinfo{person}{D. Yao}, {and}
  \bibinfo{person}{E. Bertino}.} \bibinfo{year}{2015}\natexlab{}.
\newblock \showarticletitle{Privacy-preserving detection of sensitive data
  exposure}.
\newblock \bibinfo{journal}{\emph{Transactions on Information Forensics and
  Security}} \bibinfo{volume}{10}, \bibinfo{number}{5} (\bibinfo{year}{2015}),
  \bibinfo{pages}{1092--1103}.
\newblock


\bibitem[\protect\citeauthoryear{Smarjov}{Smarjov}{2020}]%
        {Smarjov:2020}
\bibfield{author}{\bibinfo{person}{Ilja Smarjov}.}
  \bibinfo{year}{2020}\natexlab{}.
\newblock \emph{\bibinfo{title}{OWASP Secure coding practices checklist and
  training: Assessment of effectiveness in a technology company}}.
\newblock \bibinfo{thesistype}{Master's\ thesis}. \bibinfo{school}{TALLINN
  UNIVERSITY OF TECHNOLOGY}.
\newblock


\bibitem[\protect\citeauthoryear{Smith, Johnson, Murphy-Hill, Chu, and
  Lipford}{Smith et~al\mbox{.}}{2015}]%
        {Smith:2015}
\bibfield{author}{\bibinfo{person}{J. Smith}, \bibinfo{person}{B. Johnson},
  \bibinfo{person}{E. Murphy-Hill}, \bibinfo{person}{B. Chu}, {and}
  \bibinfo{person}{H. Lipford}.} \bibinfo{year}{2015}\natexlab{}.
\newblock \showarticletitle{Questions developers ask while diagnosing potential
  security vulnerabilities with static analysis}. In
  \bibinfo{booktitle}{\emph{Proceedings of the Joint Meeting on Foundations of
  Software Engineering}}. \bibinfo{pages}{248--259}.
\newblock


\bibitem[\protect\citeauthoryear{Smith, Johnson, Murphy-Hill, Chu, and
  Lipford}{Smith et~al\mbox{.}}{2018}]%
        {Smith:2018}
\bibfield{author}{\bibinfo{person}{J. Smith}, \bibinfo{person}{B. Johnson},
  \bibinfo{person}{E. Murphy-Hill}, \bibinfo{person}{B. Chu}, {and}
  \bibinfo{person}{H. Lipford}.} \bibinfo{year}{2018}\natexlab{}.
\newblock \showarticletitle{How developers diagnose potential security
  vulnerabilities with a static analysis tool}.
\newblock \bibinfo{journal}{\emph{Transactions on Software Engineering}}
  \bibinfo{volume}{45}, \bibinfo{number}{9} (\bibinfo{year}{2018}),
  \bibinfo{pages}{877--897}.
\newblock


\bibitem[\protect\citeauthoryear{Sotirov, Stevens, Appelbaum, Lenstra, Molnar,
  Osvik, and de~Weger}{Sotirov et~al\mbox{.}}{2008}]%
        {sotirov2008md5}
\bibfield{author}{\bibinfo{person}{A. Sotirov}, \bibinfo{person}{M. Stevens},
  \bibinfo{person}{J. Appelbaum}, \bibinfo{person}{A. Lenstra},
  \bibinfo{person}{D. Molnar}, \bibinfo{person}{D.~Arne Osvik}, {and}
  \bibinfo{person}{B. de Weger}.} \bibinfo{year}{2008}\natexlab{}.
\newblock \showarticletitle{MD5 considered harmful today, creating a rogue CA
  certificate}. In \bibinfo{booktitle}{\emph{Proceedings of the Annual Chaos
  Communication Congress}}.
\newblock


\bibitem[\protect\citeauthoryear{Spadini, {\c{C}}alikli, and Bacchelli}{Spadini
  et~al\mbox{.}}{2020}]%
        {spadini2020primers}
\bibfield{author}{\bibinfo{person}{D. Spadini}, \bibinfo{person}{G.
  {\c{C}}alikli}, {and} \bibinfo{person}{A. Bacchelli}.}
  \bibinfo{year}{2020}\natexlab{}.
\newblock \showarticletitle{Primers or Reminders? {T}he Effects of Existing
  Review Comments on Code Review}. In \bibinfo{booktitle}{\emph{Proceedings of
  the International Conference on Software Engineering}}.
  \bibinfo{pages}{1171--1182}.
\newblock


\bibitem[\protect\citeauthoryear{Spencer}{Spencer}{2009}]%
        {spencer2009card}
\bibfield{author}{\bibinfo{person}{D. Spencer}.}
  \bibinfo{year}{2009}\natexlab{}.
\newblock \bibinfo{booktitle}{\emph{Card sorting: Designing usable
  categories}}.
\newblock \bibinfo{publisher}{Rosenfeld Media}.
\newblock


\bibitem[\protect\citeauthoryear{Tahaei and Vaniea}{Tahaei and Vaniea}{2019}]%
        {Tahaei:2019}
\bibfield{author}{\bibinfo{person}{Mohammad Tahaei} {and} \bibinfo{person}{Kami
  Vaniea}.} \bibinfo{year}{2019}\natexlab{}.
\newblock \showarticletitle{A survey on developer-centred security}. In
  \bibinfo{booktitle}{\emph{Proceedings of the European Symposium on Security
  and Privacy Workshops}}. \bibinfo{pages}{129--138}.
\newblock


\bibitem[\protect\citeauthoryear{Thomas, Tabassum, Chu, and Lipford}{Thomas
  et~al\mbox{.}}{2018}]%
        {Thomas:2018}
\bibfield{author}{\bibinfo{person}{T. Thomas}, \bibinfo{person}{M. Tabassum},
  \bibinfo{person}{B. Chu}, {and} \bibinfo{person}{H. Lipford}.}
  \bibinfo{year}{2018}\natexlab{}.
\newblock \showarticletitle{Security during application development: An
  application security expert perspective}. In
  \bibinfo{booktitle}{\emph{Proceedings of the 2018 CHI Conference on Human
  Factors in Computing Systems}}. \bibinfo{pages}{1--12}.
\newblock


\bibitem[\protect\citeauthoryear{Thompson and Wagner}{Thompson and
  Wagner}{2017}]%
        {Thompson:2017}
\bibfield{author}{\bibinfo{person}{C. Thompson} {and} \bibinfo{person}{D.
  Wagner}.} \bibinfo{year}{2017}\natexlab{}.
\newblock \showarticletitle{A Large-Scale Study of Modern Code Review and
  Security in Open Source Projects}. In \bibinfo{booktitle}{\emph{Proceedings
  of the International Conference on Predictive Models and Data Analytics in
  Software Engineering}}. \bibinfo{pages}{83--92}.
\newblock


\bibitem[\protect\citeauthoryear{TIOBE}{TIOBE}{2020}]%
        {tiobe}
\bibfield{author}{\bibinfo{person}{TIOBE}.} \bibinfo{year}{2020}\natexlab{}.
\newblock \bibinfo{title}{{TIOBE-Index}}.
\newblock \bibinfo{howpublished}{https://www.tiobe.com/tiobe-index/}.
\newblock


\bibitem[\protect\citeauthoryear{Turpe, Kocksch, and Poller}{Turpe
  et~al\mbox{.}}{2016}]%
        {Turpe:2016}
\bibfield{author}{\bibinfo{person}{S. Turpe}, \bibinfo{person}{L. Kocksch},
  {and} \bibinfo{person}{A. Poller}.} \bibinfo{year}{2016}\natexlab{}.
\newblock \showarticletitle{Penetration Tests a Turning Point in Security
  Practices? Organizational Challenges and Implications in a Software
  Development Team}. In \bibinfo{booktitle}{\emph{Proceedings of the Symposium
  on Usable Privacy and Security}}.
\newblock


\bibitem[\protect\citeauthoryear{Vagias}{Vagias}{2006}]%
        {Vagias:2006}
\bibfield{author}{\bibinfo{person}{W. Vagias}.}
  \bibinfo{year}{2006}\natexlab{}.
\newblock \bibinfo{booktitle}{\emph{Likert-Type Scale Response Anchors}}.
\newblock \bibinfo{type}{{T}echnical {R}eport}.
\newblock


\bibitem[\protect\citeauthoryear{Weir, Rashid, and Noble}{Weir
  et~al\mbox{.}}{2016}]%
        {Weir:2016}
\bibfield{author}{\bibinfo{person}{C. Weir}, \bibinfo{person}{A. Rashid}, {and}
  \bibinfo{person}{J. Noble}.} \bibinfo{year}{2016}\natexlab{}.
\newblock \showarticletitle{How to improve the security skills of mobile app
  developers? Comparing and contrasting expert views}. In
  \bibinfo{booktitle}{\emph{Proceedings of the Symposium on Usable Privacy and
  Security}}.
\newblock


\bibitem[\protect\citeauthoryear{Weir, Rashid, and Noble}{Weir
  et~al\mbox{.}}{2017}]%
        {Weir:2017}
\bibfield{author}{\bibinfo{person}{C. Weir}, \bibinfo{person}{A. Rashid}, {and}
  \bibinfo{person}{J. Noble}.} \bibinfo{year}{2017}\natexlab{}.
\newblock \showarticletitle{I'd Like to Have an Argument, Please: Using
  Dialectic for Effective App Security}.
\newblock  (\bibinfo{year}{2017}).
\newblock


\bibitem[\protect\citeauthoryear{Wikipedia}{Wikipedia}{2021}]%
        {YahooBreach}
\bibfield{author}{\bibinfo{person}{Wikipedia}.} \bibinfo{year}{Last accessed
  August 2021}\natexlab{}.
\newblock \bibinfo{title}{Yahoo! data breaches}.
\newblock
  \bibinfo{howpublished}{https://en.wikipedia.org/wiki/Yahoo!\_data\_breaches\#Late\_2014\_breach}.
\newblock


\bibitem[\protect\citeauthoryear{Woon and Kankanhalli}{Woon and
  Kankanhalli}{2007}]%
        {woon:2007}
\bibfield{author}{\bibinfo{person}{I. Woon} {and} \bibinfo{person}{A.
  Kankanhalli}.} \bibinfo{year}{2007}\natexlab{}.
\newblock \showarticletitle{Investigation of IS professionals' intention to
  practise secure development of applications}.
\newblock \bibinfo{journal}{\emph{International Journal of Human-Computer
  Studies}} \bibinfo{volume}{65}, \bibinfo{number}{1} (\bibinfo{year}{2007}),
  \bibinfo{pages}{29--41}.
\newblock


\bibitem[\protect\citeauthoryear{Wurster and Van~Oorschot}{Wurster and
  Van~Oorschot}{2008}]%
        {Wurster:2008}
\bibfield{author}{\bibinfo{person}{G. Wurster} {and} \bibinfo{person}{P.
  Van~Oorschot}.} \bibinfo{year}{2008}\natexlab{}.
\newblock \showarticletitle{The developer is the enemy}. In
  \bibinfo{booktitle}{\emph{Proceedings of the New Security Paradigms
  Workshop}}. \bibinfo{pages}{89--97}.
\newblock


\bibitem[\protect\citeauthoryear{Xiao, Witschey, and Murphy-Hill}{Xiao
  et~al\mbox{.}}{2014}]%
        {Xiao:2014}
\bibfield{author}{\bibinfo{person}{S. Xiao}, \bibinfo{person}{J. Witschey},
  {and} \bibinfo{person}{E. Murphy-Hill}.} \bibinfo{year}{2014}\natexlab{}.
\newblock \showarticletitle{Social influences on secure development tool
  adoption: why security tools spread}. In
  \bibinfo{booktitle}{\emph{Proceedings of the Conference on Computer supported
  cooperative work and social computing}}. \bibinfo{pages}{1095--1106}.
\newblock


\bibitem[\protect\citeauthoryear{{Xie}, {Lipford}, and {Chu}}{{Xie}
  et~al\mbox{.}}{2011}]%
        {Xie:2011}
\bibfield{author}{\bibinfo{person}{J. {Xie}}, \bibinfo{person}{H.~R.
  {Lipford}}, {and} \bibinfo{person}{B. {Chu}}.}
  \bibinfo{year}{2011}\natexlab{}.
\newblock \showarticletitle{Why do programmers make security errors?}. In
  \bibinfo{booktitle}{\emph{Proceedings of the Symposium on Visual Languages
  and Human-Centric Computing}}. \bibinfo{pages}{161--164}.
\newblock


\bibitem[\protect\citeauthoryear{Zhu}{Zhu}{2016}]%
        {zhu2016software}
\bibfield{author}{\bibinfo{person}{Y. Zhu}.} \bibinfo{year}{2016}\natexlab{}.
\newblock \bibinfo{booktitle}{\emph{Software Reading Techniques}}.
\newblock \bibinfo{publisher}{Springer}.
\newblock


\end{thebibliography}

\appendix

\end{document}